\documentclass[aip, reprint, twocolumn, amsmath,amsfonts,amssymb]{revtex4-1}

\usepackage[colorlinks=true,linkcolor=blue]{hyperref}

\usepackage{graphicx}
\usepackage{amsmath,amsfonts,amssymb}
\usepackage{color}
\usepackage{bm}
\usepackage[mathscr]{eucal}
\usepackage{picture}
\usepackage{float}
\usepackage{subcaption}

\usepackage{graphicx}
\usepackage{amsmath,amsfonts,amssymb}
\usepackage{color}
\usepackage{bm}
\usepackage{threeparttable}
\usepackage[mathscr]{eucal}
\usepackage{picture}

\definecolor{grey}{rgb}{0.7,0.7,0.7}

\newcommand{\br}{\nonumber\\}
\usepackage{fancyvrb}
\usepackage{ulem}

\usepackage[T1]{fontenc}

\definecolor{brown}{RGB}{111,16,50}

\definecolor{purple}{rgb}{0.8,0.0,0.8}
\definecolor{pink}{rgb}{1.,0.5,0.5}

\newcommand{\ctext}[1]{\raise0.2ex\hbox{\textcircled{\scriptsize{#1}}}}

\begin{document}
\title{Leveraging configuration interaction singles for qualitative descriptions of ground and excited states: state-averaging, linear-response, and spin-projection}

\author{Takashi Tsuchimochi}
\email{tsuchimochi@gmail.com}
\affiliation{College of Engineering, Shibaura Institute of Technology, 3-7-5 Toyosu, Koto-ku, Tokyo 135-8548 Japan}
\affiliation{Institute for Molecular Science, 38 Nishigonaka, Myodaiji, Okazaki 444-8585 Japan}

\author{Benjamin Mokhtar}
\affiliation{Graduate School of Engineering and Science, Shibaura Institute of Technology, 3-7-5 Toyosu, Koto-ku, Tokyo 135-8548 Japan}

\begin{abstract}
While configuration interaction singles (CIS) provides a computationally efficient description of excited states, it systematically overestimates excitation energies and performs poorly for strongly correlated systems, partly due to the lack of orbital relaxation and the strong ground-state bias of Hartree–Fock orbitals. 
To address these limitations, we present a unified variational framework that extends CIS by incorporating orbital optimization in state-specific and state-averaged forms (SSCIS and SACIS), linear-response orbital relaxation via a double-CIS scheme (DCIS), and spin-symmetry breaking and restoration (ECIS). In spin-projected state-averaged formulations, standard multistate parametrizations are no longer valid because the projection operator breaks the unitary invariance of orbital rotations and induces nonorthogonal couplings among states. By formulating a rigorous state-averaged objective in the projected subspace, we derive analytic electronic gradients and Hessians and enable robust optimization using a trust-region augmented Hessian algorithm. Benchmark calculations show that spin projection alone significantly exacerbates the CIS overestimation in weakly correlated systems, whereas combining spin projection with state averaging or double-CIS corrections substantially reduces errors, particularly for Rydberg excitations. We further demonstrate that state averaging and spin projection provide complementary and essential benefits in strongly correlated regimes, as illustrated by the bond dissociation of hydrogen fluoride and nitrogen.
\end{abstract}
\maketitle

\section{Introduction}

Accurate yet computationally affordable descriptions of electronically excited states remain a long-standing challenge in quantum chemistry. High-level wave-function methods such as equation-of-motion coupled-cluster theory with singles and doubles or triples (EOM-CCSD/CCSDT)\cite{EOMCC1, EOMCC2, CC_review} and other related methods\cite{CC2, CC3} provide reliable excitation energies for a wide range of molecular systems. However, their steep computational cost severely restricts their applicability to large molecules, extended excited-state manifolds, and the exploration of excited-state potential energy surfaces. This limitation has motivated sustained efforts to develop low-cost excited-state methods that capture essential physical effects while retaining the favorable scaling and conceptual simplicity of mean-field–based approaches.

Among such methods, configuration interaction singles (CIS) represents one of the most fundamental low-cost approaches to excited states. Formally equivalent to the Tamm–Dancoff approximation (TDA) to time-dependent Hartree–Fock (TDHF), CIS provides a variationally stable and computationally efficient description of excited states in terms of single excitations from a Hartree–Fock (HF) reference. Despite these appealing features, CIS is well known to systematically overestimate excitation energies.\cite{Hirata99, Dreuw05, Subotnik2011} This deficiency originates from two fundamental limitations: the absence of dynamical correlation and, more importantly for many classes of excitations, the lack of orbital relaxation. Because excitation energies are differences between ground- and excited-state energies, their errors arise from imbalances in correlation energy between the two states. In particular, the missing differential correlation energy is typically on the order of the correlation energy associated with the electron pairs involved in the excitation. Dynamical correlation effects can often be treated perturbatively by including double excitations with respect to CIS states, leading to the widely used CIS(D) approach.\cite{CIS(D)} While treatments of dynamical correlation have a long history and are now relatively well established, the incorporation of orbital relaxation effects for excited states has attracted renewed interest in recent years.\cite{Gilbert2008, Kowalczyk2013, Levi2020, Hait2021, SSCI, SSCC, Tuckman2023, COOX, Selenius2024}

Because HF orbitals are optimized solely for the ground state, CIS excited states are inherently biased toward the ground-state electronic structure, leading to a systematic imbalance between ground and excited states. This bias is particularly severe for excitations that involve substantial redistribution of electron density, such as charge-transfer\cite{Subotnik2011, Liu2014} and core-excited states,\cite{Core1, Core2, NOCI, ESMF_core} as well as for near-degenerate electronic states.\cite{Tsuchimochi15, Tsuchimochi15B} 
In addition, CIS, or more broadly, single-reference linear-response methods including TDDFT, also fail qualitatively in strongly correlated regimes, such as bond dissociation and conical intersections,\cite{Failure_in_ConicalIntersection1, Failure_in_ConicalIntersection2, Failure_in_ConicalIntersection3} where the single-reference picture breaks down.

Orbital relaxation therefore plays a central role in improving CIS-based excited-state methods. One rigorous way to incorporate orbital relaxation is full state-specific orbital optimization for each excited state. This philosophy underlies excited-state mean-field theory (ESMF), introduced by Neuscamman and co-workers,\cite{ESMF1, ESMF2, ESMF3, ESMF4} which can be viewed as a state-specific, orbital-optimized CIS (SSCIS) method. In ESMF, or equivalently SSCIS, orbitals and CI coefficients are optimized simultaneously for a targeted excited state, yielding substantial improvements in excitation energies through balanced orbital relaxation especially when combined with perturbative treatments of dynamical correlation.\cite{ESMF1, ESMP2, ESMP2_2} However, this approach introduces a highly nonlinear optimization problem that is variationally unstable and potentially difficult to converge in practice. Moreover, the resulting excited states are state-specific and nonorthogonal, complicating the description of multiple excited states and the evaluation of transition properties.

An alternative strategy is to incorporate orbital relaxation effects within a linear-response–like framework while retaining the orthogonality and conceptual simplicity of CIS. A pioneering work by Subotnik and co-workers introduced a single-shot perturbative orbital relaxation correction to CIS using the HF Hessian as an approximation to the CIS Hessian.\cite{Subotnik2011} While this approach improved charge-transfer excitation energies, it was found to be insufficient in general.\cite{Liu2013,Liu2014} More recently, double CIS (DCIS) proposed by one of us provides a systematic framework,\cite{tsuchimochi_double_2024} in which CIS is effectively performed on top of CIS to incorporate first-order orbital relaxation effects. DCIS can be interpreted as a linear-response correction within the TDA framework and has been shown to systematically lower CIS excitation energies for charge-transfer states, achieving improved accuracy through error cancellation between ground and excited states, at modest additional computational cost. Importantly, this approach preserves orthonormality between states, enabling straightforward evaluation of transition properties.

To mitigate the strong ground-state bias inherent in HF-based excited-state methods, state averaging provides yet another route. By optimizing orbitals with respect to an average energy over several electronic states, state-averaged approaches reduce the preferential stabilization of the ground state and yield a more balanced description of excited states. State averaging has a long history in multiconfigurational self-consistent field (MCSCF) theories\cite{SACASSCF1, SACASSCF2}, and also has been combined with other methods,\cite{SAREKS, SAResHF} particularly for photochemical applications and near-degenerate electronic structures. When applied to CIS-based methods, state-averaged CIS (SACIS) should provide a natural framework for introducing orbital relaxation ``on average'' across multiple states, as a variationally stable alternative to state-specific orbital optimization. However, whether and to what extent this strategy mitigates the systematic overestimation of CIS excitation energies remains to be systematically assessed.

Strong static correlation presents an additional challenge for CIS-based excited-state methods. In such regimes, the HF reference becomes qualitatively incorrect, leading to severe failures of CIS. Spin-unrestricted formulations can partially alleviate this problem by allowing symmetry breaking, but they introduce spin contamination that complicates the interpretation of excited states. Spin-projection techniques\cite{Lowdin55,Scuseria11}, although neither size-consistent nor size-extensive,\cite{Castano86,Jimenez12} offer a systematic remedy by restoring spin symmetry while retaining the flexibility of broken-symmetry determinants. Spin-extended CIS (ECIS),\cite{Tsuchimochi15,Tsuchimochi15B} built upon a spin-projected unrestricted HF (SUHF) reference (also known as projected HF),\cite{Jimenez12} provides a natural extension of CIS for strongly correlated systems and has been shown to yield qualitatively correct excitation energies in prototypical near-degenerate cases.\cite{Tsuchimochi15, Tsuchimochi15B} However, its adequacy across a broad range of excitations and correlation regimes remains largely unexplored.

It is also important to note that, despite these conceptual advances, orbital-optimization in CIS and ECIS is expected to pose significant numerical challenges just like MCSCF. The simultaneous optimization of orbitals and CI coefficients leads to strongly coupled, nonlinear equations, for which convergence is far from trivial. Conventional schemes based on (quasi-)Newton methods\cite{MCSCF_Newton1, MCSCF_Newton2} or DIIS (direct inversion of the iterative subspace)\cite{Pulay80,Pulay82} would often suffer from slow convergence, numerical instability, or convergence to saddle points, particularly when the initial trial states are far from convergence due to the presence of near-degeneracies or multiple competing solutions. Robust optimization algorithms are therefore essential for making these methods practically viable. One promising approach is the trust-region augmented Hessian (TRAH) method,\cite{Helmich-Paris21, Helmich-Paris22} which provides enhanced robustness against such instabilities and enables reliable convergence for both state-specific and state-averaged formulations. 

In this paper, we present a unified framework for state-specific and state-averaged orbital-optimized CIS and ECIS, including their linear-response extensions based on DCIS. We derive analytic gradients and Hessians of SACIS and SAECIS with respect to orbital rotations and CI coefficients, formulate efficient and robust optimization strategies using TRAH, and systematically assess the performance of these methods for both weakly and strongly correlated systems. Through benchmark calculations and detailed analyses of convergence behavior and potential energy surfaces, we aim to clarify the respective roles of orbital relaxation, state averaging, and spin projection in low-cost excited-state methods, and to delineate their regimes of applicability.

\section{Theory}
In this work, the occupied and virtual spin orbitals are indicated by $i,j,k,l$ and $a,b,c,d$, respectively. In addition, we will indicate the $\beta$ orbitals by a bar on each index, when we discuss spatial orbitals. 
To simplify the aufbau ($|\Phi_0\rangle$) and non-aufbau, singly-excited determinants ($|\Phi_i^a\rangle$), we use $\mu$ to indicate the excitation manifold, i.e., $|\Phi_\mu\rangle = \hat E_\mu |\Phi_0\rangle \in \{\Phi_0, \Phi_i^a\}$ for generalized CIS. Capital letters $I, J, K, L$ are reserved for electronic states.

\subsection{State-specific (single-state) orbital-optimized ECIS}
\subsubsection{Energy, gradients, and Hessian}\label{sec:Theory_A1}
We begin with the state-specific (single-state) formulation. Because the projected formalism reduces straightforwardly to standard (unprojected) CIS by removing the projection operator $\hat P$, we introduce spin projection from the outset and later recover the non-projected expressions as a special case. 

An unperturbed CIS state is given by
\begin{align}
|0\rangle = \hat C | \Phi_0\rangle = \sum_{\mu} c_\mu |\Phi_\mu\rangle 
\label{eq:0}
\end{align}
where 
\begin{align}
    \hat C = \sum_\mu c_\mu \hat E_\mu
\end{align}
Spin-restricted CIS is obtained by imposing $c_{ai} = c_{\bar a\bar i}$. While this restriction enforces a pure singlet manifold, it can also reduce the flexibility needed to describe near-degeneracy and static correlation effects. Allowing spin-unrestricted single excitations and orbitals can provide the required flexibility, particularly for (near-)degenerate systems, but it typically results in  spin-symmetry breaking when the $\alpha$ and $\beta$ orbitals are optimized independently. When spin contamination is substantial, the interpretation of excited states becomes difficult---a serious drawback in photochemical applications, where distinguishing spin manifolds is essential (e.g., for studying intersystem crossing and internal conversion). 

In spin-extended CIS (ECIS), one can remove the spin contamination in Eq.~(\ref{eq:0}) to fully realize its potential, by employing spin-projection through the projection operator $\hat{P}$. In the following, we primarily focus on the spin-extended formulation including $\hat{P}$; however, the non-projected results (without the ``E'') can be readily obtained by setting $\hat{P} \rightarrow 1$. 

Here, our aim lies in constructing a fully variational ECIS wave function for a single state, whose energy is stationary with respect to both orbital rotations and CI amplitudes. Clearly, this can be achieved by introducing a set of variational parameters ${\bm \lambda} = ({^{\rm o}\bm\lambda}, {^{\rm c}\bm\lambda})^\top$, where the superscripts ${\rm o}$ and ${\rm c}$ indicate the orbital-rotation and CI subspaces, respectively. Although the presence of $\hat P$ somewhat complicates the parametrization, the state-specific case remains considerably simpler than the state-averaged formulation discussed later.

First, the orbital rotation part should be parametrized by the exponential of single excitations as usual,
\begin{align}
   {}^{\rm o}{\hat \lambda} &= \sum_{ai}{^{\rm o}}\lambda_{ai} \hat E^-_{ai}\label{eq:lambda_o}
\end{align}
with 
\begin{align}
    \hat E^-_{pq} &= \hat E_{pq } - \hat E_{qp}\\
    \hat E_{pq} & = a_p^\dag a_q
 \end{align}
Here, one only needs to take into account orbital rotations between the occupied and virtual spaces as in Eq.~(\ref{eq:lambda_o}), because an ECIS wave function is invariant with respect to occupied-occupied and virtual-virtual rotations. 

In passing, we should emphasize that for full optimization orbital rotations must be unrestricted and applied to the broken-symmetry CIS state $|0\rangle$ {\it before} performing spin-projection $\hat P$, as $\hat P e^{-{}^{\rm o}{\hat \lambda}} |0\rangle$. This corresponds to the so-called ``variation-after-projection'' scheme, which allows for full optimization of the ansatz. Conversely, if orbital rotations were to be applied {\it after} projection, i.e., $e^{-{}^{\rm o}{\hat \lambda}} \hat P|0\rangle$, the operator $^{\rm o}\hat \lambda$ would need to be spin-adapted, making it impossible to vary the degree of symmetry breaking in $|0\rangle$ and $|\Phi_\mu\rangle$. 

Second, the variational parametrization for the CI space is achieved by adding
\begin{align}
    |{^{\rm c}}{\bm\lambda}\rangle = {}^{\rm c}\hat \lambda |\Phi_0\rangle = \sum_{\mu} {^{\rm c}}\lambda_\mu|\Phi_\mu\rangle
\end{align}
with 
\begin{align}
    {}^{\rm c}\hat \lambda = \sum_\mu {^{\rm c}}\lambda_\mu \hat E_\mu
\end{align}
which takes the same form as Eq.(\ref{eq:0}) but with the CI coefficients {\bf c} replaced by $^{\rm c}\lambda_\mu$. It is advantageous to keep the CI perturbation ${^{\rm c}}{\bm\lambda}$ within the complement of the reference subspace created by {\bf c}, especially in the presence of both spin-projection and orbital rotation; namely, we require $\hat P e^{-{}^{\rm o}{\hat \lambda}} |0\rangle \perp \hat P e^{-{}^{\rm o}{\hat \lambda}} |{}^{\rm c} {\bm\lambda} \rangle$. This condition would be easily satisfied independent of ${}^{\rm o}\bm\lambda$ if spin-projection is not present, because of the unitarity of $e^{-{}^{\rm o}{\hat \lambda}}$. 
However, as the spin-projection operator $\hat P$ does not commute with (spin-unrestricted) orbital rotations, we generally have
\begin{align}
    e^{{}^{\rm o}\hat \lambda}\,\hat P\, e^{-{}^{\rm o}\hat \lambda} \neq 1,
\end{align}
and one needs to explicitly remove the CI-unperturbed ECIS wave function $\hat Pe^{-{}^{\rm o}{\hat \lambda}}|0\rangle$ from $\hat P e^{-{}^{\rm o}{\hat \lambda}} |{}^{\rm c} {\bm\lambda} \rangle$ while still allowing for orbital rotation. This can be accomplished by introducing the following projector:
\begin{align}
        \hat Q &= 1 - \frac{\hat P e^{-{}^{\rm o}{\hat \lambda}}|0\rangle \langle 0| e^{{}^{\rm o}{\hat \lambda}}\hat P}{\langle 0| e^{{}^{\rm o}{\hat \lambda}}\hat P e^{-{}^{\rm o}{\hat \lambda}}|0\rangle}
\end{align}
This definition of $\hat Q$ correctly accounts for the change in the normalization (denominator) and ensures a clear separation between the parameter spaces $\{{}^{\rm o}\lambda_{ai}\}$ and $\{{}^{\rm c}\lambda_{\mu}\}$.

These requirements lead us to the following parameterization of the state-specific, orbital-optimized ECIS (SSECIS),
\begin{align}
|\tilde 0[{\bm\lambda}]	\rangle = \hat P e^{-{}^{\rm o}{\hat\lambda}}\left(|0\rangle 	 + e^{{}^{\rm o}{\hat \lambda}}\hat Q e^{-{}^{\rm o}{\hat \lambda}}|{^{\rm c}\bm\lambda}\rangle \right)\label{eq:SS0tilde}
\end{align}
which can be optimized by finding the stationary point of the following energy expectation value:
\begin{align}
    E[{\bm\lambda}] = \frac{\left\langle\tilde 0[{\bm\lambda}]	\middle|\hat H\middle|\tilde 0[{\bm\lambda}]	\right\rangle}{\left\langle\tilde 0[{\bm\lambda}]	\middle|\tilde 0[{\bm\lambda}]	\right\rangle } \label{eq:E_SSECIS}
\end{align}
It is important to note that the norm of $|\tilde 0\rangle$ is generally not preserved, i.e., $\langle \tilde 0[{\bm\lambda}] | \tilde 0[{\bm\lambda}]	\rangle \ne 1$.
Nevertheless, it is convenient to choose the unperturbed state $|\tilde 0[{\bf 0}]\rangle \equiv \hat P|0\rangle$ to be normalized, $\langle 0|\hat P |0\rangle = 1$, simplifying the derivation. For example,
\begin{align}
    E[{\bf 0}] = \langle 0|\hat H \hat P|0\rangle \equiv E_0
\end{align}

A stationary point of Eq.~(\ref{eq:E_SSECIS}) is where the first derivatives of the energy with respect to $^{\rm o}\lambda_{ai}$ and $^{\rm c}\lambda_{\mu}$ around ${\bm \lambda} = {\bf 0}$ are both sufficiently small. They are, respectively,
\begin{align}
     {^{\rm o}}g_{ai} &= \frac{\partial E[{\bm\lambda}]}{\partial {^{\rm o}}\lambda_{ai}}\Big|_{{\bm \lambda} = {\bf 0}} = \langle 0|\left[\hat E^-_{ai}, (\hat H - E_0)\hat P\right]|0\rangle\\
    {^{\rm c}}g_{\mu} &= \frac{\partial E[{\bm\lambda}]}{\partial {^{\rm c}}\lambda_{\mu}}\Big|_{{\bm \lambda} 
 = {\bf 0}} =\langle 0 |(\hat H -E_0) \hat P|\Phi_\mu\rangle
    +\langle \Phi_\mu |(\hat H -E_0) \hat P|0\rangle
\end{align}
If the spin-projection is neglected, this reduces exactly to the excited state mean-field theory\cite{ESMF1, ESMF2, ESMF3, ESMF4}. 

While gradient minimization often converges to the desired state if the initial state $|0\rangle$ is sufficiently close to the target, it is still advantageous to incorporate the second derivatives especially for the ground state optimization when the HF state is not qualitatively accurate. 
Recently, one of us has reported the electronic Hessian for the (non-projected) CIS energy\cite{tsuchimochi_double_2024}. Here, we generalize that result to ECIS by including the spin-projection operator:
\begin{widetext}
\begin{align}
    {^{\rm oo}}H_{ai,bj} & = \left.
     \frac{\partial^2 E}{\partial {^{\rm o}}\lambda_{ai}\partial {^{\rm o}}\lambda_{bj}}\right|_{{\bm \lambda} = {\bf 0}}
     \br
     &=
    \frac{1}{2}\langle 0|\left[\hat E^-_{ai}, \left[\hat E^-_{bj}, (\hat H - E_0)\hat P\right]\right]|0\rangle
   +\frac{1}{2}\langle 0|\left[\hat E^-_{bj}, \left[\hat E^-_{ai}, (\hat H - E_0)\hat P\right]\right]|0\rangle
- {^{\rm o}}g_{ai} \langle 0|\left[\hat E^-_{bj}, \hat P\right]|0\rangle 
    - {^{\rm o}}g_{bj} \langle 0|\left[\hat E^-_{ai}, \hat P\right]|0\rangle
 \\
  {^{\rm oc}}H_{ai, \mu} & = \left.
     \frac{\partial^2 E}{\partial {^{\rm o}}\lambda_{ai}\partial {^{\rm c}}\lambda_{\mu}}\right|_{{\bm \lambda} = {\bf 0}}
    \br
    &=\langle 0| \left[\hat E^-_{ai}, (\hat H - E_0) \hat P  \right]|\Phi_\mu\rangle 
    +  \langle \Phi_\mu| \left[\hat E^-_{ai}, (\hat H - E_0) \hat P  \right]|0\rangle 
-{^{\rm o}}g_{ai} \left(\langle 0|\hat P |\Phi_\mu\rangle  +\langle \Phi_\mu|\hat P |0\rangle\right)
    - {^{\rm c}}g_{\mu}\langle 0|\left[\hat E^-_{ai}, \hat P\right]|0\rangle \\
{^{\rm cc}}H_{\mu\nu} &  = \left.
     \frac{\partial^2 E}{\partial {^{\rm c}}\lambda_{\mu}\partial {^{\rm c}}\lambda_{\nu}}\right|_{{\bm \lambda} = {\bf 0}}
    \br
   & =\langle \Phi_\mu| \hat Q(\hat H -E_0) \hat P \hat Q|\Phi_\nu\rangle +\langle \Phi_\nu|\hat Q(\hat H -E_0)  \hat P \hat Q|\Phi_\mu\rangle \br
\end{align}
\end{widetext}
For real orbitals, these can be recast as
\begin{align}
      {^{\rm oo}}H_{ai,bj}  &= {\mathscr P}(ai){\mathscr P}(bj) (2\;{^{\rm oo}}A_{ai,bj} +   {^{\rm oo}}B_{ai,bj} + {^{\rm oo}}B_{bj,ai})
    \br
    & + 2{^{\rm o}}g_{ai} {\mathscr P}(bj)P_{bj} 
     + 2{^{\rm o}}g_{bj} {\mathscr P}(ai)P_{ai}\\
    {^{\rm oc}}H_{ai,\mu}    &= -2{\mathscr P} (ai) ( {^{\rm oc}}A_{ai,\mu} +  {^{\rm oc}}B_{ai,\mu} - {^{\rm c}}g_\mu P_{ai})
    \br&
    -2{^{\rm o}}g_{ai}\langle 0|\hat P|\Phi_\mu\rangle \\
       {^{\rm cc}}H_{\mu\nu} &=2\; {^{\rm cc}} A_{\mu\nu}
   -{^{\rm c}}g_{\mu}\langle 0|\hat P|\Phi_\nu\rangle
   -{^{\rm c}}g_{\nu}\langle 0|\hat P|\Phi_\mu\rangle
\end{align}
where ${\mathscr P}(pq)$ is the anti-symmetrizer between $p$ and $q$, and 
\begin{align}
{^{\rm oo}}A_{ai,bj} &= \langle 0|\hat E_{ia} (H-E_0)\hat P \hat E_{bj}|0\rangle,\\
{^{\rm oc}}A_{ai,\mu} &= \langle 0|\hat E_{ia} (\hat H - E_0)\hat P |\Phi_\mu\rangle,\\
{^{\rm cc}}A_{\mu\nu} &= \langle \Phi_\mu| (\hat H - E_0)\hat P |\Phi_\nu\rangle,\\
{^{\rm oo}}B_{ai,bj} &= \langle 0|(\hat H -E_0)\hat P \hat E_{ai} \hat E_{bj}|0\rangle,\\
{^{\rm oc}}B_{ai,\mu} &= \langle 0| (\hat H - E_0)\hat P \hat E_{ai}|\Phi_\mu\rangle,\\
P_{pq} &= \langle 0| \hat E_{qp} \hat P  |0\rangle.
\end{align}

Note that ${^{\rm oo}}A_{ai,bj}$ and ${^{\rm oo}}B_{ai,bj}$ resemble the familiar matrices that appear in the Hartree-Fock electronic Hessian. However, they contain $\hat P$ and the de-excitation parts such as ${^{\rm oo}}A_{ia,jb}$ and  ${^{\rm oo}}B_{ia,jb}$ that are nonzero, as opposed to Hartree-Fock. Especially, the density matrix-like quantity $P_{pq}$ is asymmetric in the presence of $\hat P$.
The explicit working equations to evaluate these terms are given in the Supplemental Information.

Using the above results, quadratic convergence methods such as the Newton method and the trust-region augmented Hessian method can be formulated. We defer the discussion to Sec.\ref{sec:TRAH} where the latter method is discussed using more generalized state-averaged ECIS.

\subsubsection{Treatment of other excited states: double CIS}

One of the main disadvantages of SSCIS/SSECIS is that orbitals are optimized solely with respect to a particular CIS/ECIS state, which significantly deteriorates the description of other CIS/ECIS states. For instance, when the ground state is optimized using generalized CIS/ECIS, the other solutions, which are orthogonal to each other and thus correspond to excited states, exhibit excessively high energies, leading to a substantial overestimation of excitation energies. In the ground state SSECIS, the overestimation is even more pronounced. To avoid this issue, one could perform state-specific calculations independently for different states of interest in the same spirit as $\Delta$SCF; however, the use of different orbitals makes the optimized states non-orthogonal to each other, complicating the computation of inter-state quantities such as transition dipole moments. 

To address this challenge, we exploit the recently proposed double CIS scheme.\cite{tsuchimochi_double_2024} DCIS aims to incorporate orbital relaxation effects in CIS by performing an additional CIS. Since DCIS involves the diagonalization of its Hamiltonian matrix, the resulting states are orthogonal to one another, in a similar way to CIS.

The wave function ansatz for DCIS is represented by the two-layered CIS:
\begin{align}
    |\Psi_{\rm DCIS}\rangle &= \hat D |0\rangle = \hat D \hat C|\Phi_0\rangle\label{eq:DCIS}\\
    \hat D &= \sum_{pq} d_{pq} \hat E_{pq}
\end{align}
This scheme corresponds to a linear-response of the CIS state $|0\rangle$ within the Tamm-Dancoff approximation and treats both single excitations and de-excitations with respect to it. Note that we can utilize the orbital-optimized CIS for $|0\rangle$. 

As described in Ref.\onlinecite{tsuchimochi_double_2024}, the parameterization in Eq.~(\ref{eq:DCIS}) suffers from numerous redundancies, introducing numerical challenges. However, by reformulating it orthogonal, the same variational space can be constructed by the following parametrization without redundancies:
\begin{align}
     |\Psi_{\rm DCIS}\rangle 
     &= \sum_{ai}d_{ai} \hat E_{ai}|0\rangle + \bar d_0 |\Phi_0\rangle + \sum_{ai}\bar d_{ai} |\Phi_i^a\rangle
\end{align}
As is expected, it is straightforward to extend it to spin-extended DCIS (EDCIS) simply by adding $\hat P$. 
The variational condition for DCIS/EDCIS is given by the following equations:
\begin{align}
 \langle 0|\hat E_{ia} (\hat H - E_{\rm EDCIS})\hat P |\Psi_{\rm DCIS}\rangle &= 0\\
 \langle \Phi_\mu| (\hat H - E_{\rm EDCIS})\hat P |\Psi_{\rm DCIS}\rangle &= 0
\end{align}
Again, for the standard DCIS, one can remove $\hat P$ from the equations.
The EDCIS Hamiltonian matrix---the second derivative of the energy expectation value with respect to $\{d_{ai}, \bar d_0, \bar d_{ai}\}$---parallels the electronic Hessian matrix {\bf H}. Namely, 
\begin{align}
    &\sum_{bj}(^{\rm oo}A_{ai,bj} - \omega\;{^{\rm oo}}S_{ai,bj}) d_{bj} \br
    &+\sum_{\nu}(^{\rm oc}A_{ai,\nu} - \omega\;{^{\rm oc}}S_{ai,\nu}) \bar d_{\nu} = 0\\
    &\sum_{bj}(^{\rm co}A_{\mu,bj} - \omega\;{^{\rm co}}S_{\mu,bj}) d_{bj} \br
    &+\sum_{\nu}(^{\rm cc}A_{\mu\nu} - \omega\;{^{\rm cc}}S_{\mu\nu}) \bar d_{\nu} = 0\\ 
\end{align}
where $\omega = E_{\rm EDCIS} - E_{0}$ is the relaxation or excitation energy, and ${\bf S}$ is the overlap matrix,
\begin{align}
{^{\rm oo}}S_{ai,bj}&=\langle 0|\hat E_{ia}\hat P \hat E_{bj}|0\rangle\\
{^{\rm oc}}S_{ai,\mu}&=\langle 0|E_{ia}\hat P |\Phi_\mu\rangle\\
{^{\rm cc}} S_{\mu\nu}&=\langle \Phi_\mu| \hat P |\Phi_\nu\rangle \end{align}
Then, $\omega$ can be obtained by solving the generalized eigenvalue problem
\begin{align}
    \begin{pmatrix}
        {^{\rm oo}}{\bf A} & {^{\rm oc}}{\bf A}\\
        {^{\rm co}}{\bf A} & {^{\rm cc}}{\bf A}
    \end{pmatrix}
    \begin{pmatrix}
        {\bf d}\\ \bar {\bf d}
    \end{pmatrix}
    = \omega    \begin{pmatrix}
        {^{\rm oo}}{\bf S} & {^{\rm oc}}{\bf S}\\
        {^{\rm co}}{\bf S} & {^{\rm cc}}{\bf S}
    \end{pmatrix}
    \begin{pmatrix}
        {\bf d}\\ \bar {\bf d}
    \end{pmatrix}
    \label{eq:DCISH}
\end{align}

Using the ground state of orbital-optimized SSCIS or SSECIS for $|0\rangle$, the lowest eigenvalue of DCIS and EDCIS is zero ($\omega= 0$) as a consequence of the generalized Brillouin condition. This indicates that no additional orbital relaxation effect occurs for the ground state. In contrast, the higher-lying DCIS/EDCIS solutions correspond to the relaxed excited states relative to the SSECIS ground state, since EDCIS is, in essence, a linear-response method that performs CIS on top of SSECIS. This property is therefore expected to lead to improved excitation energies when EDCIS is combined with an SSECIS reference. 

Finally, the structure of Eq.~\ref{eq:DCISH} strongly suggests a close connection to a time-dependent formulation involving the {\bf B} matrix, where DCIS is obtained as its Tamm-Dancoff approximation. Although such a relationship between DCIS and time-dependent CIS appears natural, we do not pursue this direction further in the present work.

\subsection{State-averaged (multistate) formalism}
\subsubsection{Energy, gradients, and Hessian}
An alternative method for achieving a balanced description of ground and excited states is state-averaged (SA) orbital optimization. This approach is particularly beneficial for systems exhibiting energetic quasi-degeneracy among multiple states, such as conical intersections, as it can treat them on an equal footing. Consequently, the SA optimization has become a standard protocol in the study of photochemistry.

Since both CIS and ECIS diagonalize a truncated Hamiltonian and can thus determine multiple states simultaneously, it is conceptually straightforward to formulate their SA variants, which minimize the averaged energy of $n$ states $\{|0_I\rangle;\; I=1,\cdots,n\}$ or a weighted sum of their energies. 
For convenience, we choose the CI basis in which both the Hamiltonian and the overlap of the subspace are diagonal:
\begin{align}
   \langle 0_I|\hat H \hat P |0_J\rangle &= E_{0,I}\delta_{IJ}\label{eq:ortho1},\\
  \langle 0_I | \hat P |0_J\rangle &= \delta_{IJ}\label{eq:ortho2}
\end{align}
where $E_{0,I}$ is the energy of $\hat P |0_I\rangle$.

It should first be noted that, for multistate calculations of standard MCSCF (without $\hat P$)  such as SA-CASSCF, one frequently employs the following parameterization:\cite{Helgaker_book}
\begin{align}
    |\tilde 0_I\rangle = e^{-^{\rm o}\hat \lambda}e^{-\hat {\mathscr R}}|0_I\rangle
\end{align}
where $\hat {\mathscr R}$ is a linear combination of state-transfer operators,
\begin{align}
    \hat {\mathscr R} = \sum_{K>J, J\le n} {\mathscr R}_{KJ} (|0_K\rangle \langle 0_J| - |0_J\rangle \langle 0_K|)
\end{align}
This formulation ensures the orthogonality.

However, it is important to point out that the presence of the projection operator $\hat P$ no longer allows for this convenient parameterization, as broken-symmetry states $\{|0_I\rangle\}$ are not orthonormal to each other by themselves: they only become so when projected by $\hat P$, as shown in Eq.(\ref{eq:ortho2}). If $\hat P$ is explicitly incorporated into $\hat {\mathscr R}$, so that each of $|0_K\rangle$ and $|0_J\rangle$ is projected and orthonormal, then the situation becomes significantly more complicated. This is because, as described in Section \ref{sec:Theory_A1}, the orbital rotation $e^{-^{\rm o}\hat \lambda}$ must be applied to $|0_I\rangle$ but not to $\hat P |0_I\rangle$, and these two requirements cannot be satisfied simultaneously.

Due to this issue, we instead parameterize the $I$th ECIS state as
\begin{align}
|\tilde 0_I[{^{\rm o}}{\bm\lambda},{^{\rm c}}{\bm\lambda}^I]	\rangle = \hat P e^{-^{\rm o}\hat \lambda}|0_I\rangle 	 + \hat P \hat {\cal Q}[{^{\rm o}\bm\lambda}]  e^{-^{\rm o}\hat \lambda}|{^{\rm c}}{\bm\lambda}^I\rangle \label{eq:SA0tilde}
\end{align}
analogously to the SS scheme, Eq.~(\ref{eq:SS0tilde}).
To ensure the orthogonality between different states under perturbation of CI coefficients (but not orbital rotation), we introduce the following generalized projection operator 
\begin{align}
\hat {\cal Q} = 1 - \sum_{IJ} \hat P e^{-{}^{\rm o}\hat \lambda} |0_I\rangle ({\bf N}[{}^{\rm o}{\bm\lambda}]^{-1})_{IJ} \langle 0_J| e^{{}^{\rm o}\hat \lambda}\hat P
\end{align}
where
\begin{align}
    ({\bf N}[{}^{\rm o}{\bm\lambda}])_{IJ} = \langle 0_I| e^{{}^{\rm o}\hat \lambda} \hat P e^{-{}^{\rm o}\hat \lambda} |0_J\rangle
\end{align}
is the state overlap Eq.(\ref{eq:ortho2}) generalized with orbital rotation, and therefore is dependent on ${}^{\rm o}{\bm \lambda}$.
Therefore, $\hat {\mathcal Q}$ explicitly projects out the residual CI space that is orthogonal to the CI-unperturbed states $\{\hat P e^{-^{\rm o}\hat \lambda}|0_I\rangle\}$. However, as is the case in the SS formalism, even the parametrization of Eq.~(\ref{eq:SA0tilde}) is still not sufficient to preserve the norm and orthogonality of the ECIS states under the perturbation of ${^{\rm o}\bm\lambda}$. In fact, it violates the orthogonality between different orbital-perturbed ECI states when ${^{\rm o}\bm\lambda} \ne {\bf 0}$,
\begin{align}
    \langle \tilde 0_I |\tilde 0_J\rangle \ne \delta_{IJ}
\end{align}
due to the first term in Eq.~(\ref{eq:SA0tilde}). 
This clearly indicates that the orbital perturbation in $|\tilde 0_J\rangle$ implicitly affects the energy of other states $|\tilde 0_I\rangle$, $E_I$, even though the latter is not an explicit function of the former, and vice versa. Therefore, the simple average of each energy expectation value, $\frac{1}{n}\sum_I \langle \tilde 0_I|\hat H |\tilde 0_I\rangle/\langle \tilde 0_I | \tilde 0_I\rangle$, is not a proper function to optimize for state-averaged ECIS. In other words, the energy expectation value does not represent the true orthogonal energy $E_I$, and it is necessary to properly account for the couplings between $|\tilde 0_I\rangle$ and $|\tilde 0_J\rangle$,
\begin{align}
    {\mathcal H}_{IJ}[{^{\rm o}}{\bm\lambda},{^{\rm c}}{\bm\lambda}^I,{^{\rm c}}{\bm\lambda}^J] &= \langle \tilde 0_I[{^{\rm o}}{\bm\lambda},{^{\rm c}}{\bm\lambda}^I] | \hat H |\tilde 0_J[{^{\rm o}}{\bm\lambda},{^{\rm c}}{\bm\lambda}^J] \rangle 
    \\ 
 {\mathcal N}_{IJ}[{^{\rm o}}{\bm\lambda},{^{\rm c}}{\bm\lambda}^I,{^{\rm c}}{\bm\lambda}^J] &= \langle \tilde 0[{^{\rm o}}{\bm\lambda},{^{\rm c}}{\bm\lambda}^I] |\tilde 0[{^{\rm o}}{\bm\lambda},{^{\rm c}}{\bm\lambda}^J] \rangle
  \end{align}
Note that in standard non-projected methods such as SA-CIS, orthogonality is guaranteed, i.e., ${\mathcal N}_{IJ} = \delta_{IJ}$, because of the unitarity of $e^{-^{\rm o}\hat \lambda}$. The emergence of the non-orthogonal property in spin-projection methods is attributed to the fact that $e^{^{\rm o}\hat \lambda} \hat P e^{-^{\rm o}\hat \lambda} \ne \hat P$.
One could avoid this non-orthogonal issue by introducing an appropriate projector in our ansatz Eq.(\ref{eq:SA0tilde}) to ensure orthogonality, but this would introduce significant complications.
Nevertheless, we can simplify the formulation if we focus on the {\it averaged} energy rather than a weighted sum of energies. 

The energy $E_I$ can be defined only after explicit orthogonalization, which involves solving the generalized eigenvalue problem
\begin{align}
    {\bm{\mathcal H}}{\bf V} = {\bm{\mathcal N}}{\bf V}{\bf E} \label{eq:HVSVE}
\end{align}
where $\bm{\mathcal H}$ and $\bm{\mathcal N}$ are the Hamiltonian matrix and overlap matrix in the subspace $|\tilde 0_I\rangle;\;(I=1,\cdots,n)$, {\bf V} is the eigenvector, and {\bf E} is the diagonal matrix composed of $E_I$. This allows us to express the averaged energy as
\begin{align}
    E_{\rm ave} = \frac{{\rm Tr}[{\bm{\mathcal {N}}}^{-1}{\bm {\mathcal H}}]}{n}
\end{align}
where we should stress that the numerator corresponds to the sum of the energies $\sum_I E_I = {\rm Tr}[{\bf E}]$, as evident from Eq.(\ref{eq:HVSVE}).
This way, $E_{\rm ave}$ can be written as an explicit function of the parameter set $
    \bm\lambda = (
        {^{o}\bm\lambda}, {^{c}\bm\lambda}^1, {^{c}\bm\lambda}^2,\cdots,{^{c}\bm\lambda}^{n})^\top
    $.

The first derivatives are quite similar to those of the state-specific scheme. It is straightforward to find
\begin{align}
    {^{\rm o}}g_{ai} &\equiv \frac{\partial E_{\rm ave}}{\partial \;^{\rm o}\lambda_{ai}}\Bigg|_{{\bm\lambda} ={\bf 0}} 
    = \frac{1}{n}\sum_{I}^{n} {^{\rm o}}g_{ai}^{I^*I}\\
    {^{\rm c}}g^I_\mu &\equiv \frac{\partial E_{\rm ave}}{\partial \;{^{\rm c}\lambda^I_\mu}}\Bigg|_{{\bm\lambda} ={\bf 0}} 
   \br
   & =\frac{1}{n} \left(\langle 0_I |(\hat H -E_{0,I}) \hat P|\Phi_\mu\rangle
    +\langle \Phi_\mu |(\hat H -E_{0,I}) \hat P|0_I\rangle\right)
    \label{eq:cgI}
\end{align}
where 
\begin{align}
    {^{\rm o}}g_{ai}^{I^*K} &= \langle 0_I | \left[\hat E_{ai}^-, (\hat H - E_{0,I})\hat P\right]|0_K\rangle
\end{align}
is the gradient-like coupling term. Here, the star on $I$ indicates the reference energy is that for $I$ (i.e., $E_{0,I}$).

The second derivatives with respect to ${\bm\lambda}$ are important not only for orbital optimization but also for geometry optimization (nuclear gradients), as will be discussed in a separate paper.\cite{tsuchimochi2026}
Again, in real orbitals,
  \begin{align}
            {^{\rm oo}}H_{ai,bj}  &= \frac{1}{n}\sum_I^n \Big({\mathscr P}(ai){\mathscr P}(bj) \big(2\;{^{\rm oo}}A_{ai,bj}^I +   {^{\rm oo}}B_{ai,bj}^I \br
            &+ {^{\rm oo}}B_{bj,ai}^I\big)
   + \sum_K^n \big({^{\rm o}}g_{ai}^{I^*K} {\mathscr P}(bj) (P_{bj}^{IK} +  P_{bj}^{KI})
      \br
    & + {^{\rm o}}g_{bj}^{I^*K} {\mathscr P}(ai)(P_{ai}^{IK} +  P_{ai}^{KI})\big)\Big)
\\
    {^{\rm oc}}H^I_{ai,\mu}    &= - \frac{2}{n} {\mathscr P} (ai) ( {^{\rm oc}}A_   {ai,\mu}^I +  {^{\rm oc}}B_{ai,\mu}^I - {^{\rm c}}g_\mu^I P^I_{ai})
    \br&
    -\frac{2}{n}\sum_{K}^n {^{\rm o}}g_{ai}^{I*K}  \langle 0_K |\hat P| \Phi_\mu\rangle \\
       {^{\rm cc}}H_{\mu\nu}^{IJ} &= \frac{\delta_{IJ}}{n} \Bigg(2{^{\rm cc}}A^I_{\mu\nu}
    -\sum_K^{n} \Big( {^{\rm c}}g^{K}_\mu  \langle 0_K|\hat P|\Phi_\nu\rangle
\br&    + {^{\rm c}}g^{K}_\nu \langle 0_K |\hat P|\Phi_\mu\rangle
\br&     +2( E_{0,K} - E_{0,I}) \langle \Phi_\mu|\hat P|0_K\rangle \langle 0_K|\hat P|\Phi_\nu\rangle\Big)
 \Bigg)
\end{align}
where ${^{\rm oo}}A^I_{ai,bj}$, etc., are similarly defined for the $I$th state, and $P_{ai}^{IK}$ is the transition density matrix between $I$ and $K$,
\begin{align}
    P_{pq}^{IK} = \langle 0_I| \hat E_{qp} \hat P |0_K\rangle
\end{align}
It is clear that in this basis there is no direct coupling between $I$th and $J$th states through their CI coefficients, and they are indirectly coupled with each other through the $\rm oc$-block. In other words, the CI part of the Hessian is block-diagonal. The structure of the Hessian matrix is therefore as follows:
\begin{align}
   {\bf H} &= \begin{pmatrix}
         {^{\rm oo}}{\bf H} &  {^{\rm oc}}{\bf H}^1 &  {^{\rm oc}}{\bf H}^2 & \cdots  &{^{\rm oc}}{\bf H}^{n}\\
          {^{\rm co}}{\bf H}^1 &  {^{\rm cc}}{\bf H}^{11} & {\bf 0} & \cdots & {\bf 0}\\
       {^{\rm co}}{\bf H}^2  & {\bf 0} &  {^{\rm cc}}{\bf H}^{22}& \cdots & {\bf 0}\\
        \vdots & \vdots & \vdots &\ddots & \vdots\\
    {^{\rm co}}{\bf H}^{n}  & {\bf 0}  & {\bf 0}&\cdots & {^{\rm cc}}{\bf H}^{nn} \\
    \end{pmatrix}
\end{align}
One caveat is that this Hessian contains a few zero eigenvalues because of the redundant parameterization introduced in Eq.~(\ref{eq:SA0tilde}).

We have summarized the detailed derivation and the working equations in the Supplemental Information.

As a useful interpretation of the orbital-optimization framework, it is worth mentioning that, for an RHF reference, SACIS can be interpreted as a special case of state-averaged RASSCF.\cite{RASSCF,Malmqvist90} Specifically, the RAS1 space corresponds to the full occupied orbital space, RAS2 is empty, and RAS3 contains the full virtual orbital space, with the restriction that at most one hole in RAS1 and one particle in RAS3 are allowed. In this sense, SACIS is formally equivalent to a state-averaged RASSCF description restricted to single excitations. Despite this formal connection, the SACIS implementation differs substantially from a conventional RASSCF algorithm. In particular, SACIS avoids molecular-orbital transformations of two-electron integrals by performing all contractions in the atomic-orbital basis, and it eliminates redundant orbital rotations within the occupied and virtual subspaces, which do not affect CIS energies. These features lead to a considerably more efficient implementation.

\begin{table*}[t]
\centering
\caption{Number of Fock-like matrices $N_{\rm Fock}$ required for orbital optimization and CI iterations in SACIS and SAECIS. SAECIS further incurs a prefactor of $N_g$ due to grid integration.}
\label{tb:cost}
\begin{tabular}{lcccc}
\hline\hline
Method & Iteration type & Operation & SACIS & SAECIS \\
\hline
TRAH & Macro &${^{\rm o}}{\bf g}^{I^*K}$, ${^{\rm c}}{\bf g}^I$, {\bf F} 
     & $\dfrac{n(n+1)}{2} + n + 1$ 
     & $\dfrac{n(n+1)}{2} + 2n + 1$ \\[6pt]
TRAH & Micro & ${\bf H}{\bm\lambda}$
     & $4n + 1$ 
     & $7n + 1$ \\[6pt]
DIIS & Macro & ${^{\rm o}}{\bf g}^{I^*I}, {\bf F}$
     & $n + 1$ 
     & $2n + 1$ \\[6pt]
DIIS & Micro & ${^{\rm c}}{\bf g}^I$ 
     & $n$ 
     & $n$ \\
\hline\hline
\end{tabular}
\end{table*}

\section{SCF algorithm}
In this section, we discuss two algorithms for orbital optimization in SACIS and SAECIS. 
\subsection{Newton method and trust-region augmented Hessian method}\label{sec:TRAH}
In the Newton method, the parameters are updated by solving
\begin{align}
           {\bf H} 
        {\bm \lambda}
=
    -{\bf g}
\end{align}
where
\begin{align}
    {\bf g} = ({^{\rm o}}{\bf g}, {^{\rm c}}{\bf g}^1, \cdots , {^{\rm c}}{\bf g}^n)^\top
\end{align}
However, this update is often not optimal when the states are (on average) far from the minimum, where the energy functional is not well approximated by a quadratic form.\cite{Helgaker_book} Instead, the trust-region augmented Hessian (TRAH) algorithm diagonalizes the augmented Hessian,
\begin{align}
   \begin{pmatrix}
        0 & \alpha {\bf g}^\top \\
        \alpha{\bf g} & {\bf H}\\
    \end{pmatrix}
    \begin{pmatrix}
        1\\ {\bm\lambda}(\alpha) 
    \end{pmatrix}
    =
    \mu
    \begin{pmatrix}
        1\\ {\bm\lambda}(\alpha) 
    \end{pmatrix}
    \label{eq:TRAH}
\end{align}
where $\mu$ is the lowest eigenvalue of the augmented Hessian (AH) matrix and acts as a level shift. The scaling factor $\alpha$ confines the update ${\bm\lambda}$ within a trust radius $h$, thereby avoiding abrupt changes:
\begin{align}
    \frac{1}{\alpha^2} \| {\bm \lambda}\|^2 \le h^2
    \label{eq:TR}
\end{align}
Further details of the TRAH method can be found in the literature.\cite{Host08, Helmich-Paris21, Helmich-Paris22} 

The TRAH algorithm has a two-level structure. In each macro iteration, the gradient {\bf g} and Hessian {\bf H} of the given CIS/ECIS states are constructed, and the AH is diagonalized using an iterative eigensolver such as the Davidson method.\cite{Davidson75} The iterative scheme of Davidson helps dynamically determine an appropriate value of $\alpha$ satisfying Eq.(\ref{eq:TR}). Upon convergence of the micro iterations, the resulting ${\bm\lambda}$ is used to update the MO coefficient matrix
as 
\begin{align}
    {\bf C}_{\rm new} = {\bf C}_{\rm old}\exp({-{\bm\kappa}})
\\
    {\bm \kappa} = \begin{pmatrix}
        {\bf 0} & -{^{\rm o}}{\bm \lambda}^\top\\
        {^{\rm o}}{\bm \lambda}&{\bf 0}
    \end{pmatrix}
\end{align}
while the CI coefficients are updated by adding $ {^{\rm c}}{\bm\lambda}^I$. As noted above, in SAECIS, each projected state $\hat P|0_I\rangle$ must subsequently be re-orthonormalized. 

We note that an exact solution of Eq.(\ref{eq:TRAH}) in each macro iteration is not required, particularly during the initial macro cycles, since the update ${\bm \lambda}$ is only a quadratic approximation. Therefore, we relax the convergence criterion for the micro iterations to $\|{\bf g}_{\rm micro}\|\le \beta \|{\bf g}\|$, where ${\bf g}_{\rm micro}$ denotes the residual vector of the micro iterations and $0< \beta < 1$ is a hyper parameter. In this work, we set $\beta = 0.2$. 

During the SCF optimization, the trust radius $h$ is adaptively updated to ensure both robustness and efficiency. Specifically, we employ Fletcher's algorithm to control $h$ at each macro step\cite{Fletcher}. Although detailed procedures are described in other works\cite{Helmich-Paris21, Helmich-Paris22}, we briefly review the algorithm here in order to introduce our modification.

In Fletcher's algorithm, two energy differences are computed:
\begin{align}
    \Delta E_{\rm actu} = E_{\rm new} - E_{\rm old}\\
    \Delta E_{\rm pred} = {\bf g}^\top {\bm\lambda} + \frac{1}{2} {\bm\lambda}^\top {\bf H} {\bm\lambda}
\end{align}
where $\Delta E_{\rm actu}$ is the actual energy change and $\Delta E_{\rm pred}$ is the predicted change based on the quadratic model. Their ratio $r = \Delta E_{\rm actu}/\Delta E_{\rm pred}$ is used to update $h$ as follows:
\begin{itemize}
    \item $r>0.75$: scale $h$ by 1.2.
    \item $0.25 < r\le 0.75$: keep $h$ unchanged.
    \item $0 \le r \le 0.25$: scale $h$ by 0.7.
    \item $r<0$: scale $h$ by 0.7, reject the orbital update and repeat the micro iterations with the reduced $h$.
\end{itemize}

In the present work, however, even when the energy increases after the orbital update (i.e., $r < 0$), we do not reject the update. Such energy increases frequently occur when the current solution is near a saddle point, indicating the presence of lower-energy solutions. At a saddle point, without the trust radius constraint (i.e., $\alpha = 1$), the update $\bm\lambda$ would become excessively large; however, the trust-region restriction effectively limits the step size and guides the optimization toward lower-energy states. A similar strategy has been adopted in the quadratically convergent algorithm for SUHF,\cite{Uejima20} motivated by the same considerations.

Finally, we consider the computational cost of the TRAH algorithm in SACIS and SAECIS for $n$ states.
Because the augmented Hessian in Eq.~(\ref{eq:TRAH}) is diagonalized using the Davidson method, the rate-limiting steps are the construction of the gradient vector ${\bf g}$ (in each macro iteration) and the evaluation of the sigma vector ${\bf H}{\bm\lambda}$.
Nevertheless, the formal computational scaling of these operations remains $O(N^4)$, where $N$ is the number of orbitals.
This scaling is dominated by the construction of several Fock-like matrices, which involve contractions between AO integrals and pseudo-density matrices (see the Supporting Information). 

In each macro iteration, the standard Fock matrix {\bf F} for the reference $|\Phi_0\rangle$ is constructed in the current orbital basis. The evaluation of the CI gradients ${^{\rm c}}{\bf g}^I$ and the (transition) orbital gradients ${^{\rm o}}{\bf g}^{I^*K}$ requires $n$ and $n(n+1)/2$ Fock-like matrices, respectively. For SAECIS, there is an additional overhead of $n$. Since Eq.~(\ref{eq:cgI}) is symmetric (assuming real orbitals), and thus only one of the two equivalent contractions is required for the sigma vector of standard CIS and ECIS. In contrast, the TRAH algorithm requires explicit contributions from both terms; therefore, SAECIS needs two Fock-like matrices per state rather than one. Finally, ECIS requires numerical grid integration for spin projection, increasing the cost by a factor of $N_g$, where $N_g$ is the number of grid points. 

For the sigma vector in each micro iteration, $4n+1$ ($7n+1$ for SAECIS) additional Fock-like matrices are constructed.
Therefore, although the formal scaling remains $O(N^4)$, orbital optimization entails a significantly larger prefactor than single-shot CIS and ECIS calculations, as expected.

\subsection{DIIS combined with effective Fock}
Although quadratic convergence is desirable for orbital optimization, it is also of interest to explore the use of DIIS as an alternative convergence acceleration scheme for SACIS and SAECIS.
In conventional SCF procedures, DIIS updates the Fock matrix ${\bm{\mathcal F}}$ by minimizing its off-diagonal elements, which vanish at self-consistency according to the Brillouin theorem; the updated orbitals are then obtained by diagonalizing ${\bm{\mathcal F}}$.
For SACIS/SAECIS orbital optimization, while the off-diagonal elements are clearly identified with the residual vector,
i.e., ${\mathcal F}_{ai} \equiv {^{\rm o}g}_{ai}$,
there is no obvious definition for the diagonal blocks ${\mathcal F}_{ij}$ and ${\mathcal F}_{ab}$.

To circumvent this issue, we define an effective Fock matrix as
\begin{align}
    {\mathcal F}_{ij} &= F_{ij},\\
    {\mathcal F}_{ab} &= F_{ab} + \delta_{ab}\,\epsilon_{\rm LS},
\end{align}
where $F_{pq}$ is the standard Fock matrix constructed from the current HF-like determinant $|\Phi_0\rangle$, and $\epsilon_{\rm LS}$ is a level-shift parameter introduced to stabilize the orbital update.
Diagonalization of the effective Fock matrix,
\begin{align}
    {\bm{\mathcal F}}{\bf U} = {\bf U}{\bm{\epsilon}},
\end{align}
yields a unitary transformation that mixes occupied and virtual orbitals, and the MO coefficients are updated as
\begin{align}
    {\bf C}_{\rm new} = {\bf C}_{\rm old}{\bf U}.
\end{align}
This approach has been taken in the SCF of symmetry-projected HF including SUHF.\cite{Scuseria11, Jimenez12}

The application of DIIS to SACIS and SAECIS then closely follows the standard DIIS protocol:
the effective Fock matrices and corresponding error vectors from several iterations are transformed into a common orbital representation (the AO basis), and an optimal linear combination of the Fock matrices is determined by minimizing the norm of the error vector.

The computational cost of this approach is expected to be lower than that of the TRAH algorithm, since constructing the effective Fock matrix requires only $n+1$ Fock-like matrices for SACIS ($n$ for the orbital gradients of each CIS state and one for the HF-like reference state) and $2n+1$ for SAECIS per macro iteration.
For the micro iterations (standard CIS and ECIS calculations), the Davidson algorithm requires $n$ Fock-like matrices to construct the sigma vector.
The number of Fock-like matrices, $N_{\rm Fock}$, required for each method is summarized in Table~\ref{tb:cost}.

However, it is well known that DIIS is generally ineffective when the initial orbitals are far from convergence.
In addition, the effective Fock matrix is well defined only for converged CIS states, for which ${^{\rm c}g_\mu} = 0$ and the orbital gradient ${^{\rm o}g}_{ai}$ is available.
Consequently, the use of DIIS implies a two-step optimization strategy for SACIS and SAECIS, in which the orbital and CI-coefficient optimizations are decoupled.

\subsection{Point-group symmetry}\label{sec:PG}
In spin-projection approaches, point-group symmetry is not explicitly enforced, and the converged wave function is therefore not guaranteed to retain the correct point-group symmetry, since spin-unrestricted methods allow the breaking of spatial symmetry in addition to spin symmetry.
To explicitly account for point-group symmetry within the framework of symmetry breaking and restoration, a point-group symmetry projection may be desirable.\cite{Jimenez12, Jimenez13,Uejima17} However, from our experiences in many cases this is not needed; spin-projection can resurrect the lost spatial symmetry if not all. This means each of the contaminated spin states possesses a different point-group symmetry than the one the designated spin state has. Therefore, we will not perform the point-group symmetry projection, but simply assign each excited state to the dominant point-group irreducible representation (irrep).

\begin{figure*}
    \includegraphics[width=120mm]{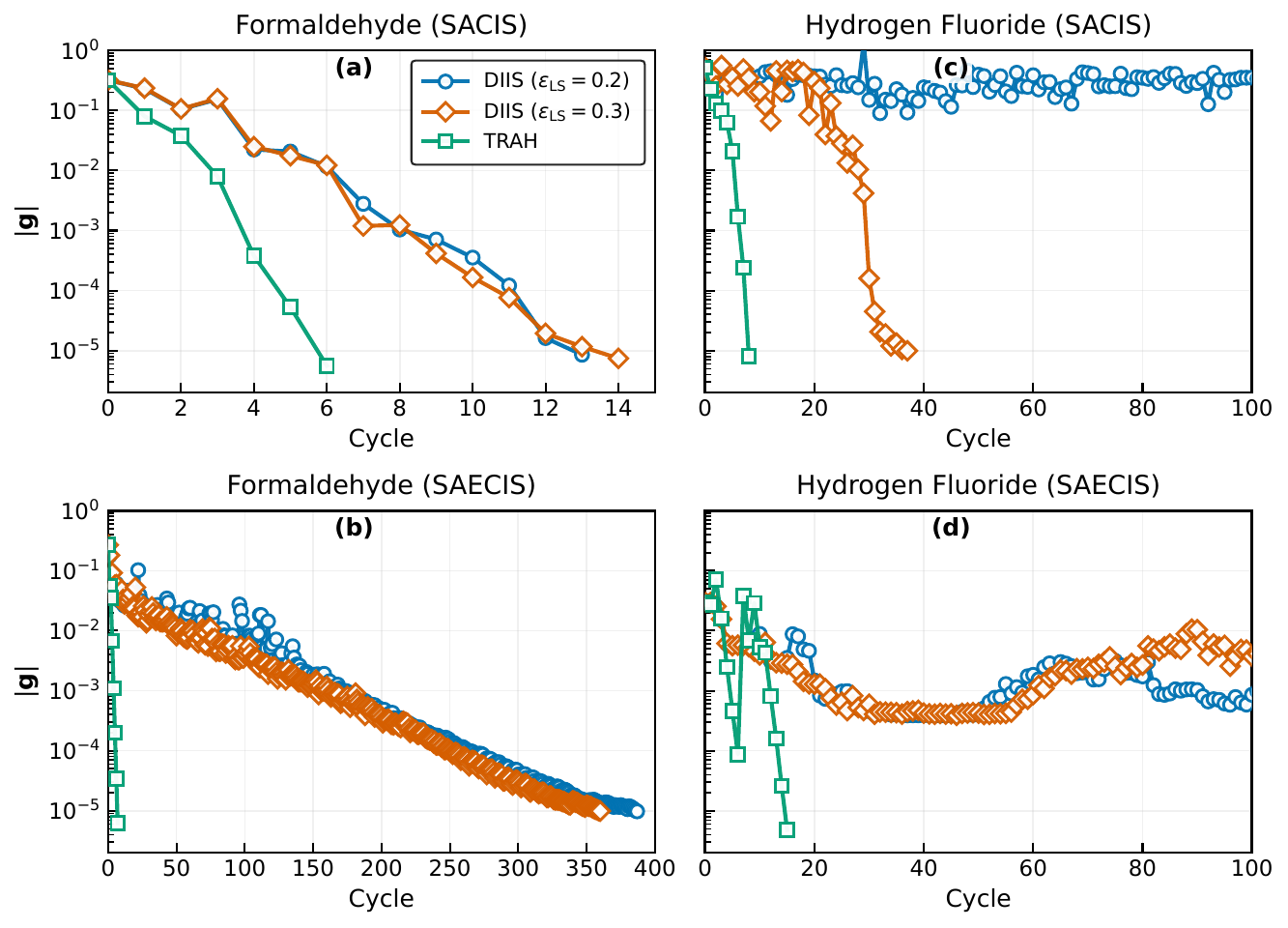}
    \caption{Comparison between convergence behaviors of DIIS and TRAH. (a) Formaldehyde with SACIS. (b) Formaldehyde with SAECIS. (c) Hydrogen fluoride with SACIS. (d) Hydrogen fluoride with SAECIS.}\label{fig:conv}
\end{figure*}

\begin{figure*}
    \includegraphics[width=120mm]{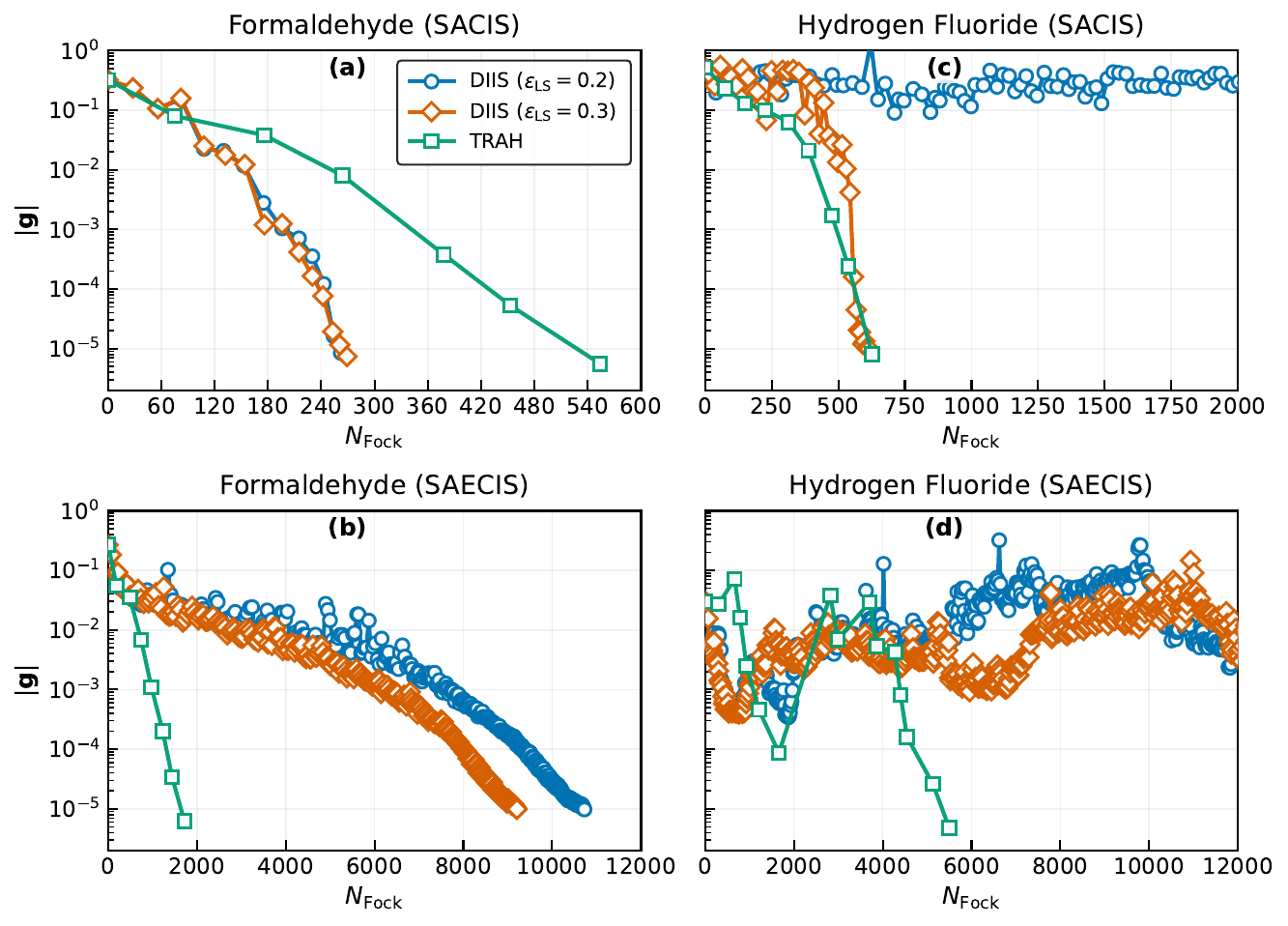}
    \caption{Same as Figure~\ref{fig:conv} but as a function of $N_{\rm Fock}$.}\label{fig:conv_Nfock}
\end{figure*}

To assign the dominant irrep of each state $|\Psi_I\rangle$, we evaluate the overlap 
\begin{align}
    \omega_\Gamma = \langle \Psi_I| \hat P_{\Gamma} | \Psi_I\rangle
\end{align}
where $\hat P_{\Gamma}$ is the projector to the irrep $\Gamma$,
\begin{align}
    \hat P_{\Gamma} = \sum_{\hat R} \chi_{\Gamma}(\hat R) \hat R
\end{align}
Here, $\hat R$ are the symmetry operations of the point group under consideration, and $\chi_\Gamma(\hat R)$ are the characters. Noting that $\sum_\Gamma \omega_\Gamma = 1$, $\omega_\Gamma$ stands for the weight of the component of the irrep $\Gamma$ in $|\Psi_I\rangle$.  In all our calculations below, $\omega_\Gamma$ is more than 90 \% (in most cases 100\%) for a certain $\Gamma$ and almost zero for others, by which we properly assign the irrep to facilitate the direct comparisons with the reference excitations.

\section{Results}
\subsection{Computational details}
All CIS, ECIS, SACIS, SAECIS, DCIS, and EDCIS calculations were performed using the {\sc Gellan} suite of programs.\cite{Gellan} For complete-active-space SCF (CASSCF) and related methods, we used ORCA.\cite{orca}
The 6-31G basis set was used for hydrogen fluoride and nitrogen, while aug-cc-pVDZ was employed for all other molecules. 

For the TRAH method, the initial value of the trust radius $h$ was set to 0.3.  In the effective Fock approach, simultaneous updates of multiple CIS/ECIS states were carried out using the block-Davidson algorithm with diagonal preconditioning, where the diagonal was approximated without explicit two-electron integrals. The level shift $\epsilon_{\rm LS}$ applied to the effective Fock matrix was set to 0.2--0.3. Convergence was assumed when the norm of the corresponding gradient fell below $10^{-5}$. For DIIS acceleration, we used eight error vectors for SACIS and five for SAECIS. The latter converges much more slowly than SACIS, and the DIIS inversion often suffers from linear dependence.

For SACIS and DCIS, spin-restricted calculations were enforced so that all computed states were singlets. In contrast, for spin-projected methods, all parameters were optimized spin-unrestrictedly, and four grid points were used for numerical integration. To perform SACIS and SAECIS calculations, we first converged the corresponding CIS and ECIS states using restricted HF and SUHF orbitals, respectively, and subsequently used them as initial guesses. 
 
In state-averaged schemes, the number of states included in the averaging must be chosen appropriately. In practice, it is typically sufficient to include the lowest few states that may become nearly degenerate along the potential energy surface of interest. We find that the qualitative features of the results are not overly sensitive to moderate variations in the number of averaged states, provided that the relevant low-lying states are included. 
For the benchmark calculations in Section~\ref{sec:benchmark}, we state-averaged over a fixed
manifold of $n$ states that includes prescribed numbers of roots from each irrep in the benchmark set
(including the ground state). 
In all methods, however, point-group symmetry constraints were not enforced during the orbital
optimization.
Therefore, for the SA optimization of multiple states with designated symmetry, we employed DIIS and,
in each macro iteration, first converged more than $n$ CIS/ECIS roots in the micro step.
From these converged roots, we then identified and selected the lowest-energy solutions with the
desired irreps following the protocol described in Section~\ref{sec:PG}, and used only this selected
set to construct the SA orbital gradient for the subsequent DIIS update.

\subsection{Convergence behavior}
We first examine the convergence performance of the TRAH method and DIIS for SACIS and SAECIS. The test systems are formaldehyde at its equilibrium geometry and hydrogen fluoride at a stretched bond length of $R_{\rm H-F} = 3.0~\mathrm{\AA}$. In all cases, the three lowest states ($n=3$), including the ground state, were optimized. For formaldehyde, these correspond to the $n \rightarrow \pi^*$ and $n \rightarrow 3s$ excited states, while for hydrogen fluoride they correspond to doubly degenerate $\pi \rightarrow \sigma^*$ states. Both TRAH and DIIS yield the same energies upon successful convergence.

Because the HF ground state of formaldehyde at equilibrium is stable, the two lowest CIS excited states are also expected to behave well. Using these states as the initial guess for SACIS, the effective Fock approach with DIIS converges rapidly for both $\epsilon_{\rm LS}=0.2$ and 0.3, as shown in Figure~\ref{fig:conv}(a). Depending on the level shift, convergence is achieved within 13--14 iterations. In contrast, TRAH converges even faster, within six macro-iterations. However, it should be noted that TRAH requires a larger number of Fock builds ($N_{\rm Fock}=554$) than DIIS ($N_{\rm Fock}=262$ for $\epsilon_{\rm LS}=0.2$ and $269$ for $\epsilon_{\rm LS}=0.3$), because TRAH incurs additional overhead from the computation of $\sigma$ vectors during micro-iterations (see Table~\ref{tb:cost}). Nevertheless, these numbers appear to be reasonable, given that the initial CIS (starting point of SACIS) requires $N_{\rm Fock} = 86$. To enable a fairer comparison of computational cost, Figure~\ref{fig:conv_Nfock}(a) plots the gradient norm $\lVert \mathbf{g} \rVert$ as a function of $N_{\rm Fock}$. 

For SAECIS, one might anticipate behavior similar to SACIS, since the system is only weakly correlated and both SUHF and ECIS converge stably, analogous to HF and CIS. Surprisingly, however, DIIS performs very poorly in this case. As shown in Figure~\ref{fig:conv}(b), DIIS requires 350--400 iterations to reach convergence for SAECIS, corresponding to nearly 10,000 Fock builds (Figure~\ref{fig:conv_Nfock}(b)). In sharp contrast, TRAH converges within seven macro-iterations, requiring only $N_{\rm Fock}=1719$ (the initial ECIS costs $N_{\rm Fock} = 146$). These results indicate that TRAH is far more suitable than DIIS for SAECIS.

When the system becomes strongly correlated and the HF reference orbitals are qualitatively incorrect, SACIS can induce substantial orbital changes during optimization. This situation arises for stretched hydrogen fluoride. Using HF orbitals, the initial ground- and excited-state energies are $E_0=-99.6243$ and $E_1=E_2=-99.6144$~Hartree, which are significantly higher than the fully optimized SACIS energies, $E_0=-99.8585$ and $E_1=E_2=-99.8575$~Hartree. Such large orbital relaxations pose a serious challenge for DIIS. Indeed, for $\epsilon_{\rm LS}=0.2$, the effective-Fock-based orbital updates are unstable and fail to converge altogether (Figure~\ref{fig:conv}(c)). Increasing the level shift to $\epsilon_{\rm LS}=0.3$ opens the occupied--virtual gap and partially stabilizes the optimization, allowing DIIS to converge. Nevertheless, large oscillations in energies and orbitals are observed during the first ten iterations, indicating intrinsic instability, and a total of 37 iterations are required for convergence. By contrast, TRAH converges smoothly without such oscillations. Although the total number of Fock builds is comparable for the two approaches, TRAH is clearly more reliable in this strongly correlated regime.

Figures~\ref{fig:conv}(d) and \ref{fig:conv_Nfock}(d) show that DIIS applied to SAECIS exhibits similar convergence difficulties, albeit for a different underlying reason. In this case, the initial guess lies close to a saddle point on the energy surface. When started from such orbitals, DIIS becomes trapped and fails to locate the lower-energy minimum. By contrast, TRAH initially follows the saddle-point solution but successfully detects the instability in the seventh macro-iteration, as evidenced by an increase in the average energy $E_{\rm ave}$. After transitioning to the lower-energy solution, TRAH converges rapidly.

In summary, DIIS is effective in reducing computational cost for SACIS when the HF orbitals provide a reasonable reference. However, TRAH is considerably more robust and is particularly advantageous for SAECIS and strongly correlated systems. A promising practical strategy may therefore be to employ TRAH in the early stages to obtain a reliable approximate solution, and subsequently switch to DIIS to accelerate convergence at reduced computational cost.

\begin{figure*}
\includegraphics[width=\textwidth]{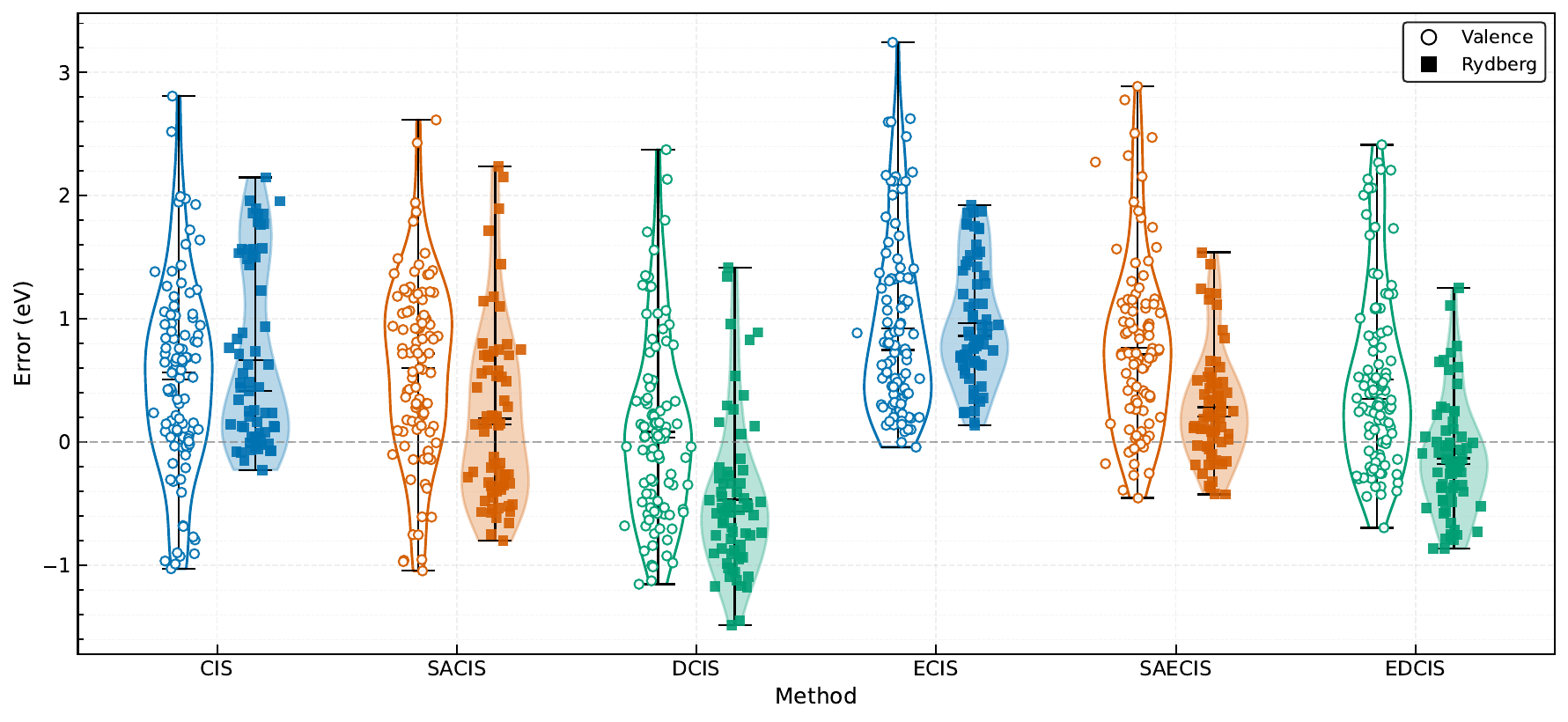}
\caption{Excitation energy error from EOM-CCSDT in eV. $\bigcirc$ and $\blacksquare$ denote valence and Rydberg excitations.}\label{fig:violin}
\end{figure*}

\subsection{Benchmark calculations for excitation energies of weakly correlated systems} \label{sec:benchmark}
To quantitatively assess the performance of the proposed CIS-based methods, we evaluate their ability to reproduce vertical excitation energies for a diverse set of valence and Rydberg excited states using the benchmark dataset of Loos \textit{et al.}~\cite{loos_mountaineering_2022}. This dataset comprises singlet excited states of various symmetries across a wide range of small organic molecules. Although none of the molecules in this benchmark exhibits strong static correlation, the availability of highly accurate CCSDT excitation energies as reference values allows for a reliable assessment of the overall accuracy of the methods in weakly correlated regimes. Among the test molecules, the two excitations for silylidene were not classified as either valence or Rydberg states and were therefore excluded from both categories.

Figure~\ref{fig:violin} summarizes the distributions of excitation energy errors for each method relative to the reference EOM-CCSDT values taken from Ref.~[\onlinecite{loos_mountaineering_2022}]. 
It is evident that most methods systematically overestimate excitation energies. The general trend is that the overestimation observed for CIS/ECIS is partially mitigated by the introduction of state averaging (SACIS/SAECIS), and further reduced by the use of the double CI scheme (DCIS/EDCIS). For valence excitations, SACIS does not offer any improvement; in fact, both the ME and MAE deteriorate from 0.51 to 0.60 eV and from 0.73 to 0.78 eV, respectively, as summarized in Table~\ref{tb:error_summary}. This behavior is not unexpected, since SACIS does not incorporate dynamical correlation effects, which constitute the dominant source of error for valence excitations. In contrast, SACIS shows a clear improvement for Rydberg excitations. The mean overestimation is significantly reduced, with the ME decreasing from 0.67 to 0.19 eV. This improvement can be attributed to the nature of Rydberg states, which typically differ substantially from the mean field (HF) electronic environment, and therefore require strong orbital relaxation to balance the description of the ground and excited states. The MAE is also reduced, albeit more modestly, from 0.69 to 0.57 eV, reflecting the continued absence of dynamical correlation. 

Similar behavior is expected for DCIS, which also seeks error cancellation between the ground and excited states by incorporating linear-response orbital relaxation effects within a CI framework. While this expectation is largely borne out, the energy lowering achieved by DCIS is more pronounced than in SACIS, leading to an overall underestimation of Rydberg excitation energies. This difference arises because SACIS optimizes orbitals by averaging over all targeted states, so that no single state is fully optimized. In contrast, DCIS employs state-specific orbital optimization for each excited state, albeit at the first-order level. Consequently, the relaxation effects in DCIS are inherently stronger than in SACIS, particularly when the number of averaged states $n$ exceeds two. Notably, the excitation energy lowering in DCIS is also observed for valence excitations, resulting in the overall improvement over CIS by 0.1 eV .

\begin{figure*}
    \includegraphics[width=150mm]{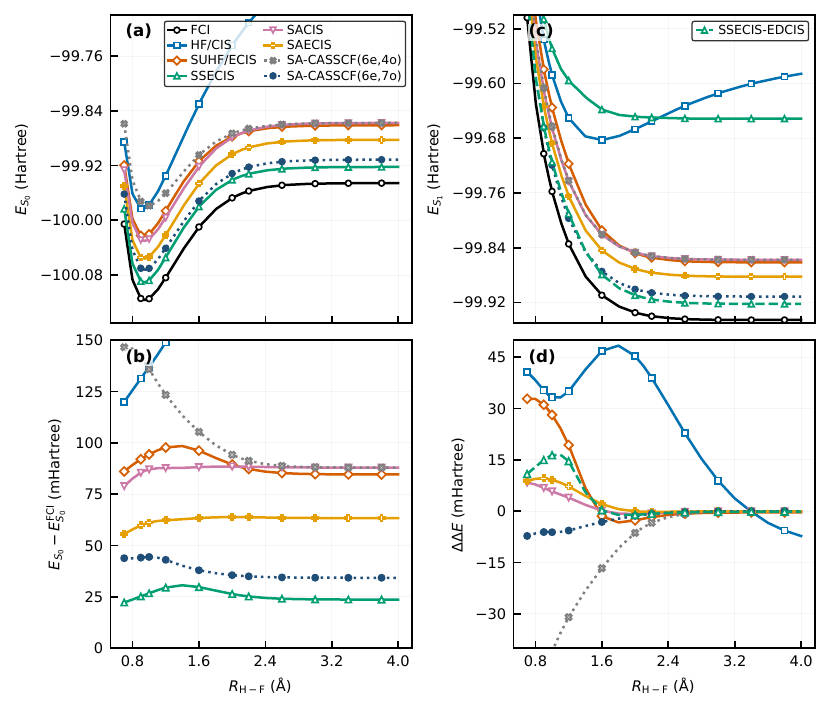}
    \caption{Potential curves of hydrogen fluoride computed by different methods. (a) Ground state $S_0$. (b) Energy error from FCI for $S_0$. (c) First excited state $S_1$. (d) Error from the FCI excitation energy.}\label{fig:FH}
\end{figure*}

One of the most striking observations from this benchmark is that ECIS substantially overestimates excitation energies for all types of excited states considered. Both the ME and MAE of ECIS reach 0.93~eV, rendering its overall performance significantly worse than that of CIS. This deficiency can likely be attributed to an overly biased description of the reference SUHF state relative to the excited states. Although SUHF and ECIS were originally developed to remedy the qualitative failures of HF and CIS in systems with near-degeneracy, their benefits are highly system dependent. Indeed, as shown previously in Fig.~\ref{fig:FH}(d), ECIS yields markedly improved excitation energies over CIS for stretched hydrogen fluoride, where strong static correlation is present. Near equilibrium geometries, however, its accuracy is comparable to that of CIS.

Taken together, these results indicate that for molecules that are only weakly correlated, spin projection provides little to no advantage over CIS and may even degrade excitation energies. It is worth noting that a time-dependent extension of SUHF was shown in the original work to substantially improve the Tamm–Dancoff ECIS results by incorporating de-excitation channels as effective orbital-relaxation effects.\cite{Tsuchimochi15} We  expect that similar improvements would also be observed for the present benchmark, in close analogy to the systematic lowering of excitation energies found in TDHF and TDDFT.\cite{Hirata99,Dreuw05,Shao16} However, time-dependent SUHF is not variational and is known to produce spurious zero eigenvalues,\cite{Tsuchimochi15} which complicates the assignment and interpretation of excited states. For this reason, we do not pursue this direction further.

To remedy the inferior performance of ECIS, both state averaging and double CI schemes are effective for improving excitation energies. As shown clearly in Figure~\ref{fig:violin} and Table~\ref{tb:error_summary}, SAECIS recovers an accuracy comparable to SACIS for valence excitations (MAE = 0.81 eV), while offering a substantially improved description for Rydberg excitations (MAE = 0.39 eV). Again, EDCIS further ameliorates the systematic overestimation and shifts excitation energies downward, reducing the ME for both valence and Rydberg states. Overall, the accuracies of SAECIS and EDCIS are comparable, with EDCIS being slightly more accurate (MAE = 0.54 eV) than SAECIS (MAE = 0.63 eV). 

Finally, we emphasize that the conclusions drawn here are specific to weakly correlated regimes. We expect that the proposed methods, particularly SAECIS and EDCIS, will perform more favorably for strongly correlated systems, where orbital relaxation and near-degeneracy effects play a more prominent role. A systematic assessment in such regimes, however, requires appropriate benchmark datasets, which are
currently unavailable and will therefore be pursued in future work. In the absence of such a dataset,
next we use the bond dissociation of hydrogen fluoride and nitrogen as stringent qualitative test cases for the proposed methods in strongly correlated regimes.

\begin{table}[htbp]
\centering
\caption{Error statistics (ME/MAE) for all, valence (V), and Rydberg (R) data sets in eV.}
\label{tb:error_summary}
\begin{tabular}{lccccccccc}
\hline
 && \multicolumn{2}{c}{All} && \multicolumn{2}{c}{V} && \multicolumn{2}{c}{R} \\
 \cline{3-4}\cline{6-7}\cline{9-10}
Method
 &\;\;& ME & MAE &\;\;
 & ME & MAE &\;\;
 & ME & MAE \\
\hline
CIS
 && 0.57 & 0.71 &
 & 0.51 & 0.73 &
 & 0.67 & 0.69 \\
SACIS
 && 0.44 & 0.69 &
 & 0.60 & 0.78 &
 & 0.19 & 0.57 \\
DCIS
 && -0.14 & 0.61 &
 & 0.08 & 0.56 &
 & -0.47 & 0.68 \\
ECIS
 && 0.93 & 0.93 &
 & 0.92 & 0.92 &
 & 0.97 & 0.97 \\
SAECIS
 && 0.56 & 0.63 &
 & 0.77 & 0.81 &
 & 0.28 & 0.39 \\
EDCIS
 && 0.24 & 0.54 &
 & 0.51 & 0.65 &
 & -0.13 & 0.39 \\
\hline
\end{tabular}
\end{table}

\subsection{Potential energy curves of hydrogen fluoride}

Because the proposed orbital-optimized approaches may be regarded as multiconfigurational SCF--type methods, it is essential to examine the quality of the resulting wave functions in the presence of strong electron correlation.

To this end, we computed the potential energy curves of hydrogen fluoride for the ground state and two $\pi\rightarrow\sigma^*$ excited states, using the same system as in the previous section. We examined CIS, ECIS, SSECIS, SACIS, and SAECIS, and compared their results with exact full CI (FCI) reference data. For comparison with conventional multireference approaches, we also performed state-averaged CASSCF calculations using both a minimal active space of (6e,4o) and a moderately enlarged active space of (6e,7o).

Figure~\ref{fig:FH}(a) shows the ground-state potential energy curves. As is well known, HF exhibits large errors as the bond is stretched. These errors arise from the lack of static correlation and clearly indicate the necessity of a multideterminantal description for bond dissociation. In contrast, SUHF incorporates static correlation through a superposition of nonorthogonal determinants and thereby recovers the qualitatively correct dissociation behavior.

Somewhat unexpectedly, SACIS is also capable of describing bond dissociation, despite the absence of explicit double excitations that would normally be required to represent the dissociation limit,
\[
\left(c_0|\phi_\sigma^2\phi_{\sigma^*}^0\rangle - c_1|\phi_\sigma^0\phi_{\sigma^*}^2\rangle\right)|\rm (Core)\rangle,
\]
where $|\rm (Core)\rangle = |\phi_{{\rm F}_{1s}}^2\phi_{{\rm F}_{2s}}^2\phi_{{\rm F}_{p_x}}^2\phi_{{\rm F}_{p_y}}^2\rangle$ denotes the closed-shell, non-bonding orbitals. Evidence of bond breaking in orbital-optimized CIS was first discussed by Burton,\cite{Burton22} who studied the hydrogen molecule. Here we show that SACIS can also describe bond dissociation in hydrogen fluoride. Inspection of the ground-state wave function at large bond distances reveals that SACIS instead favors a localized-orbital description reminiscent of valence bond theory. In this limit, the reference determinant $|\Phi_0\rangle$ corresponds to the ionic configuration H$^+$F$^-$. Single excitations from $\phi_{{\rm F}_{p_z}}$ to $\phi_{{\rm H}_{1s}}$ within CIS then yield the correct dissociation limit in the form of a charge-transfer singlet,
\begin{align}
|\Psi_{S_0}^{\rm SACIS}\rangle
= \frac{1}{\sqrt{2}}
\left(
|\phi_{{\rm H}_{1s}}^\alpha \phi_{{\rm F}_{p_z}}^\beta\rangle
+
|\phi_{{\rm H}_{1s}}^\beta \phi_{{\rm F}_{p_z}}^\alpha\rangle
\right) |\rm (Core)\rangle.
\end{align}

 Consistent with this interpretation, the potential energy curve obtained with SACIS exhibits improved behavior near dissociation. 
As shown in Figure~\ref{fig:FH}(b), the absolute ground-state energy errors of SACIS relative to FCI are comparable to those of SUHF. However, the SACIS curve is more nearly parallel to the FCI reference. This is quantified by the non-parallelity error (NPE), defined as the difference between the maximum and minimum deviations from FCI, which is reduced from 13.9~mHartree for SUHF to 9.6~mHartree for SACIS, as summarized in Table~\ref{tb:NPE}.

An even more striking feature of SACIS is its ability to correctly describe the exact degeneracy between the ground and excited states in the dissociation limit. Although this behavior was already observed at $R_{\rm H-F}=3.0$ {\AA} in the preceding section, Figs.~\ref{fig:FH}(c) and \ref{fig:FH}(d) provide a more detailed analysis. These panels show the excited-state potential energy curves $E_{S_1}$ (degenerate with $E_{S_2}$) and the deviation of the excitation energy from the FCI reference, $\Delta\Delta E$, respectively. As the bond is stretched, most methods—including SACIS—recover the vanishing excitation energy of FCI, corresponding to exact degeneracy between the states ($\Delta\Delta E = 0$). In SACIS, this degeneracy arises from charge-transfer excitations from $\phi_{{\rm F}_{p_x}}$ and $\phi_{{\rm F}_{p_y}}$ into $\phi_{{\rm H}_{1s}}$. While SACIS cannot be expected to describe all possible types of degeneracy, these results demonstrate that state averaging alone already captures the essential qualitative physics of bond dissociation and excited-state degeneracy for this system.

Having established the central role of state averaging, we next assess whether spin projection provides additional benefits for the excited-state potential energy surfaces. To this end, we examine ECIS and its state-averaged extension, SAECIS.

Near the equilibrium bond distance, where the electronic structure is only weakly correlated, ECIS offers little improvement over CIS in excitation energies. In this regime, the spin-projected SUHF reference remains strongly biased toward the closed-shell ground state, and spin projection alone cannot compensate for the absence of orbital relaxation. Consequently, ECIS excitation energies around equilibrium remain comparable to those obtained with CIS, consistent with the trends observed in the previous section.

As the bond is stretched and the system enters a strongly correlated regime, the role of spin projection becomes more pronounced. In this limit, ECIS substantially improves upon CIS by restoring the correct symmetry and near-degeneracy between configurations, thereby recovering qualitative features of the excited-state potential energy surfaces that are entirely missed by CIS. At the same time, the success of SACIS demonstrates that these qualitative features can already be captured through state averaging at the CIS level, without explicit spin projection. This indicates that while spin projection can provide a useful complementary mechanism for describing strongly correlated excited states, it is not strictly required when state averaging is properly employed.

This observation naturally motivates
the combination of spin projection with state averaging. By optimizing orbitals with respect to an average energy over multiple states, SAECIS reduces the ground-state bias inherent in ECIS and provides a more balanced description of ground and excited states along the entire potential energy surface.

As evidenced by both Figure~\ref{fig:FH} and Table~\ref{tb:NPE}, SAECIS performs very similarly to SACIS for this system in a relative sense. Although SAECIS gains additional correlation energy in absolute terms, this effect is nearly constant along the potential energy curves and therefore does not qualitatively alter their shapes. This indicates that the dominant qualitative features of the electronic structure are already captured through state averaging at the CIS level, whereas spin projection provides only additional, but comparatively modest, corrections.

While state averaging provides a balanced and robust description of multiple states, it necessarily sacrifices variational optimality for any single state. It is therefore instructive to contrast SAECIS with a purely state-specific optimization strategy, in which the orbitals are optimized exclusively for the ground state. To this end, we next examine the behavior of SSECIS.

As expected, SSECIS recovers significantly more correlation energy than SAECIS; its total energy is approximately 10~mHartree lower than that of SS-CASSCF(6e,7o) over the entire potential energy curve (not shown). However, excited states obtained na\"ively as higher-energy solutions of the underlying ECIS Hamiltonian are highly unstable (Figure~\ref{fig:FH}(c)), leading to severe overestimation of excitation energies. As a consequence, the corresponding $\Delta\Delta E$ values reach several hundred mHartree and are therefore omitted from Figure~\ref{fig:FH}(d). This failure can be attributed to the strong bias of the optimized orbitals toward ground-state energy minimization.

Introducing the DCIS scheme as a linear-response treatment on top of SSECIS largely remedies this problem. The resulting excitation energies, shown as dashed curves in Figs.~\ref{fig:FH}(c) and \ref{fig:FH}(d), exhibit substantial improvement, although the NPEs for $S_1$ and $\Delta\Delta E$ remain larger than those obtained with SACIS and SAECIS (Table~\ref{tb:NPE}).

Finally, SA-CASSCF(6e,4o) represents the minimal active space for this system, but it yields large ground-state errors, particularly near the equilibrium bond length, with an NPE of 58.6~mHartree. In contrast, the excited-state energies are reasonably described over the entire range of bond lengths, with an NPE of 7.9~mHartree. Expanding the active space to SA-CASSCF(6e,7o) significantly improves the ground-state description, resulting in overall accuracy comparable to that of SACIS and SAECIS. These results underscore the sensitivity of CASSCF to the choice of active space.

In summary, SACIS and SAECIS provide balanced and reliable descriptions of the ground- and excited-state potential energy surfaces for hydrogen fluoride. A particularly attractive feature of these methods is that they do not require a user-defined active space, unlike CASSCF, allowing them to be applied in a largely black-box manner.

\begin{table}[htbp]
\centering
\caption{Non-parallelity errors (NPE) in mHartree.}\label{tb:NPE}
\begin{tabular}{lccc}
\hline
Method & NPE($S_0$) & NPE($S_1$) & NPE($\Delta\Delta E$) \\
\hline
HF/CIS & 247.2 & 199.2 & 55.6 \\
SUHF/ECIS & 13.9 & 38.9 & 36.1 \\
SSECIS & 8.4 & 114.9\footnote{Second lowest solution of the underlying CIS Hamiltonian.}, 21.2\footnote{Linear response using the EDCIS scheme.} & 113.4$^{\rm a}$, 17.6$^{\rm b}$ \\
SACIS & 9.6 & 5.6 & 9.2 \\
SAECIS & 8.2 & 7.0 & 9.8 \\
SA-CASSCF(6e,4o) & 58.6 & 7.9 & 52.5 \\
SA-CASSCF(6e,7o) & 10.2 & 4.1 & 7.2 \\
\hline
\end{tabular}
\end{table}

\begin{figure*}
	\includegraphics[width=0.8\textwidth]{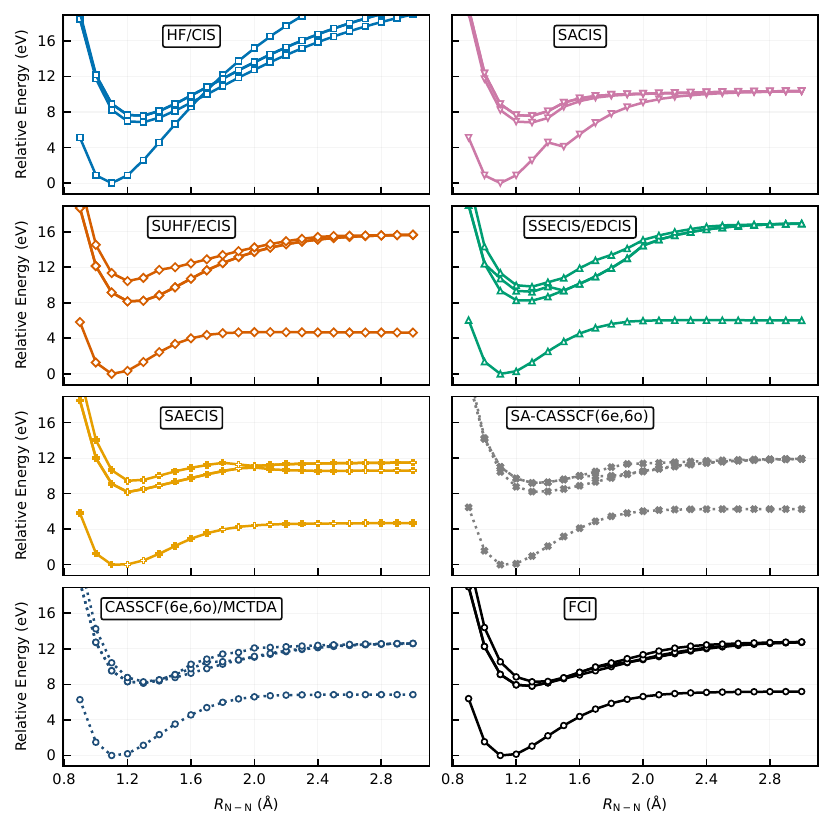}
\caption{Potential energy curves of N$_2$ for the $S_0$--$S_3$ states computed by different methods. All curves are shifted relative to the $S_0$ minimum at 1.1~\AA.}
\label{fig:N2}
\end{figure*}

\subsection{Potential energy curves of nitrogen}
As a more stringent test case for strong correlation, we consider the ground and excited states of the nitrogen molecule at varying bond distances. In addition to the methods used in the preceding sections, we also include the MCTDA approach, i.e., the linear-response formulation of CASSCF within the TDA framework.\cite{MCRPA} This allows us to examine how linear-response approaches, such as EDCIS and MCTDA, perform in describing excited states in the strongly correlated regime.

Figure \ref{fig:N2} summarizes the potential energy curves of the ground state ($S_0$) and the three lowest singlet states ($S_1$--$S_3$) regardless of irreducible representation. All curves are plotted relative to the $S_0$ energy at equilibrium ($R = 1.1$ \AA). Figure \ref{fig:N2} clearly demonstrates that SACIS is not capable of describing the triple-bond dissociation without spin-symmetry breaking; its curves are qualitatively incorrect due to the lack of static correlation, showing a behavior similar to HF/CIS. 

As expected, SSECIS shows qualitative agreement for $S_0$ with CASSCF and FCI. Its excitation energies computed with the linear-response EDCIS, however, fail to capture the strong correlation in the excited states near the dissociation limit. This behavior is similar to ECIS, where the ground state is described reasonably well but the excited states become inaccurate. This contrasts sharply with MCTDA, which shows remarkably similar results to FCI, even though these approaches are conceptually related in that a linear response is applied to the reference wave function (SSECIS or CASSCF(6e,6o)). The key difference is that MCTDA also includes the subspace spanned by all configuration variations within the active space. Consequently, the resulting excitation manifold generally contains higher excitation effects within the active space (e.g., configurations corresponding to double excitations relative to the ground-state CASSCF reference). In contrast, SSECIS does not employ an active space; therefore, the excitation manifold of EDCIS is spanned purely by single excitations from the SSECIS reference. This difference becomes particularly important in strongly correlated regimes, where single excitations alone cannot describe the multi-configurational character of excited states.

By contrast, the description provided by SAECIS is qualitatively acceptable not only for the $S_0$ state but also for the excited states across the dissociation region. Although genuine higher excitations are not explicitly included in SAECIS, their effects are partially recovered through orbital relaxation in the state-averaged framework (see the SACIS description of HF dissociation in the previous section) and through spin projection, which effectively introduces multi-configurational character into the reference wave function. This mechanism allows SAECIS to capture part of the static correlation that would otherwise require higher excitations in conventional CI-type methods. 

The SAECIS curves agree well with those of SA-CASSCF(6e,6o) at shorter bond lengths ($<2.0$ \AA), although the $S_1$--$S_3$ energies begin to split at larger separations.At the dissociation limit ($R = 3.0$ \AA), the excitation energies of $S_1$, corresponding to the $\sigma\rightarrow\sigma^*$ transition, are 5.92, 5.64, 5.74, and 5.53 eV for SAECIS, SA-CASSCF(6e,6o), MCTDA, and FCI, respectively, showing good overall agreement. In contrast, the $S_2$ and $S_3$ states, corresponding to the $\sigma\rightarrow\pi^*$ transitions, lie 0.89 eV above $S_1$ with SAECIS, whereas the other methods, including ECIS, predict these states to be nearly degenerate with $S_1$. This behavior highlights a limitation of SAECIS in capturing the full degeneracy pattern in the strongly correlated dissociation limit; nevertheless, it still provides the most reasonable overall description among the approaches proposed in this work.

\section{Conclusions}
In this work, we have developed and systematically assessed a family of low-cost CIS-based excited-state methods that incorporate spin projection, orbital relaxation, and state averaging within a unified variational framework. By formulating both state-specific and state-averaged orbital-optimized CIS and ECIS, and by extending the double-CI approach to include spin projection, we have clarified how different physical mechanisms—error cancellation, orbital relaxation, and symmetry restoration—affect excitation energies across distinct correlation regimes. In addition, we have demonstrated that robust optimization strategies are indispensable for these methods: in particular, the TRAH method, realized with the aid of analytic Hessian derived in this work, enables reliable convergence for strongly coupled nonlinear optimization problems arising in state-averaged schemes, where conventional DIIS approaches often fail or converge very slowly.

Benchmark calculations for weakly correlated molecules reveal that state averaging and linear-response–type orbital relaxation are effective in reducing the systematic overestimation of CIS excitation energies, especially for Rydberg states. In contrast, spin projection alone, as realized in ECIS, does not generally improve excitation energies in this regime and may even degrade them due to an overly biased SUHF reference. However, combining spin projection with state averaging or double-CI corrections (SAECIS and EDCIS) restores a balanced description of ground and excited states, yielding excitation energies slightly better than their non-projected counterparts. These results highlight that the usefulness of spin projection is highly regime dependent and must be complemented by appropriate orbital-relaxation mechanisms.

The importance of state-averaging becomes more evident in strongly correlated systems, as illustrated by the potential energy curves of stretched hydrogen fluoride. In this case, conventional CIS fails qualitatively due to the breakdown of the HF reference, whereas SACIS correctly captured the essential physics of near-degeneracy and static correlation. Spin-projection variants, ECIS and SAECIS, are also successful in describing excited-state potential energy surfaces in bond-breaking regimes, but SAECIS again is more robust as ECIS overestimates excitation energies around weakly correlated region (i.e, equilibrium bond distance). 

 Subsequently, the potential energy curves of the nitrogen molecule provide a more stringent test of strong correlation effects. In this case, SACIS alone is not sufficient to fully capture the static correlation associated with triple-bond dissociation. The linear-response treatment of SSECIS using EDCIS shows behavior similar to ECIS, where the multi-reference character of strongly correlated excited states cannot be described by single excitations from the ground state. In contrast, by combining spin projection with orbital optimization and state averaging, SAECIS is able to effectively mimic part of the higher-excitation effects and produces qualitatively correct potential energy curves across the dissociation region.

Naturally, the proposed methods are also expected to improve the description of charge-transfer excitations as well as core excitations, for which orbital optimization plays a crucial role. Indeed, the former are partially remedied within the DCIS framework, as demonstrated in our previous work.\cite{tsuchimochi_double_2024} For core excitations, preliminary tests indicate that DCIS significantly improves the accuracy over CIS, although a systematic benchmark study will be reported elsewhere. For the state-averaged framework considered here, however, the treatment of charge-transfer and core excitations is more challenging. These states are typically high in energy and often require state-specific, non-variational optimization rather than state averaging. Nevertheless, recent progress in related orbital-optimized excited-state approaches, such as ESMF, which has been shown to improve the description of core excitations,\cite{ESMF_core}  suggests that similar improvements may be achievable with SACIS and SAECIS when hybridized with an appropriate state-specific framework. 

We emphasize that all the methods proposed in this work remain mean-field approximations and no dynamical correlation is treated explicitly. Therefore, it will be important to incorporate dynamical correlation for these CIS variants to achieve further accuracy for ground and excited states. One natural direction is to develop perturbative correlation corrections tailored to these frameworks. Another possible route is to combine the present methods with density functional theory. In principle, this could be achieved by formulating the excitation energies within a TDDFT/TDA-like response framework based on Kohn–Sham orbitals. However, such an extension would require higher-order derivatives of the exchange–correlation functional in the orbital optimization step, which poses nontrivial implementation challenges. Nevertheless, hybrid schemes combining wave-function and density-functional ingredients may offer a promising direction for future developments.

In addition, systematic comparisons with $\Delta$SCF approaches and the evaluation of transition properties, such as transition dipole moments, would also be valuable directions for future work.

\section*{Associated content}
\subsection*{Supplemental information}
Detailed derivation of matrix elements for SAECIS and EDCIS, and simplified equations for SACIS and DCIS; 
excitation energies computed with CIS, SACIS, SAECIS, DCIS, and EDCIS.

\section*{Acknowledgments}
This work was supported by JST FOREST Program, Grant No. JPMJFR223U, and JSPS KAKENHI, Grant Nos. 25K01733 and 25K22247. B. M. acknowledges support from the SIT Scholarship of Shibaura Institute of Technology.

\section*{Notes}
The authors declare no competing financial interest.

\bibliography{SAECIS.bib}

\end{document}